# Infrared Spectroscopy of Large, Low-Albedo Asteroids: Are Ceres and Themis Archetypes or Outliers?

Andrew S. Rivkin[1], Ellen S. Howell[2], Joshua P. Emery[3]

1. Johns Hopkins University Applied Physics Laboratory
2. University of Arizona
3. University of Tennessee

Corresponding author: Andrew Rivkin (andy.rivkin@jhuapl.edu)

Key points:

- The largest low-albedo asteroids appear to be unrepresented in the meteorite collection
- Reflectance spectra in the 3-$\mu$m spectral region, and presumably compositions, similar to Ceres and Themis appear to be common in low-albedo asteroids > 200 km diameter
- The asteroid 324 Bamberga appears to have hemispherical-scale variation in its reflectance spectrum, and in places has a spectrum reminiscent of Comet 67P.

# 1 Abstract:

Low-albedo, hydrated objects dominate the list of the largest asteroids. These objects have varied spectral shapes in the 3-$\mu$m region, where diagnostic absorptions due to volatile species are found. Dawn's visit to Ceres has extended the view shaped by ground-based observing, and shown that world to be a complex one, potentially still experiencing geological activity.

We present 33 observations from 2.2-4.0 $\mu$m of eight large (D > 200 km) asteroids from the C spectral complex, with spectra inconsistent with the hydrated minerals we see in meteorites. We characterize their absorption band characteristics via polynomial and Gaussian fits to test their spectral similarity to Ceres, the asteroid 24 Themis (thought to be covered in ice frost), and the asteroid 51 Nemausa (spectrally similar to the CM meteorites). We confirm most of the observations are inconsistent with what is seen in meteorites and require additional absorbers. We find clusters in band centers that correspond to Ceres- and Themis-like spectra, but no hiatus in the distribution suitable for use to simply distinguish between them. We also find a range of band centers in the spectra that approaches what is seen on Comet 67P. Finally, variation is seen between observations for some objects, with the variation on 324 Bamberga consistent with hemispheric-level difference in composition. Given the ubiquity of objects with 3-$\mu$m spectra unlike what we see in meteorites, and the similarity of those spectra to the published spectra of Ceres and Themis, these objects appear much more to be archetypes than outliers.

# 2 Background and Motivation
## 2.1 The Largest Asteroids

There is evidence that the largest asteroids are different from the smaller ones. Collisional evolution models suggest that objects larger than 100 km or so are overwhelmingly likely to remain intact through solar system history, while those below roughly 50 km diameter are likely to be a fragment of a once larger object . Recent planetesimal creation and evolution models suggest that asteroids were "born big", going straight from mm-scale particles to objects 100 km in size or larger (Morbidelli et al. 2009).

The list of the largest asteroids is dominated by low-albedo objects. Only 6 of the 24 objects in the present-day main asteroid belt with diameters larger than 200 km have albedos higher than 0.10, and one of them is 2 Pallas (which has an albedo of 0.101 and belongs to a low-albedo taxonomic class). The remaining 18 objects are split fairly evenly in terms of Bus-DeMeo taxonomic class (DeMeo et al. 2014) between the Ch/Cgh class (5 objects), the B or Cb class (5 objects), and the C class (6 objects), with two objects in the X complex. Looking beyond the 0.5-2.5 $\mu$m region where these taxonomies are rooted, we know that diversity also exists in the 2.5-4 $\mu$m spectra of large asteroids (Rivkin et al. 2015a). Surveys (Takir et al. 2012, Rivkin 2010) and studies of individual asteroids from the ground and on site (for instance, the VIR spectrometer onboard Dawn) demonstrate a variety of band shapes in the 3-$\mu$m region, discussed in more detail in following sections.

Ceres, the largest object in the asteroid belt, has a distinctive band shape in the 3-$\mu$m region that has most recently been interpreted as due to ammoniated minerals and carbonates (De Sanctis et al. 2015). The asteroid 24 Themis has a spectral shape that has been interpreted as due to ice frost and organic materials (Rivkin and Emery 2010, Campins et al. 2010). We report observations in the 3-$\mu$m region that, together with published work, provide insight into the hydrated mineralogy of the low-albedo, C-complex objects larger than 200 km diameter in the present-day asteroid belt with spectra that are similar to Ceres and Themis.

2.2  Non-Meteoritic Hydrated Minerals on Asteroids

Ceres was the first asteroid on which hydrated minerals were detected (Lebofsky 1978, Lebofsky et al 1981), followed by Pallas and a few other large asteroids (Lebofsky 1980, Feierberg et al. 1983). It was quickly recognized that despite similarities shortward of 2.5 $\mu$m, Ceres and Pallas were unlike each other at longer wavelengths (Larson et al. 1983). Laboratory measurements of meteorites found that carbonaceous chondrite spectra were similar to Pallas rather than Ceres (Hiroi et al. 1996, Sato et al. 1997, Rivkin et al. 2003). Due to technical limitations at the time, initial measurements of asteroids in these wavelengths were focused on determining the presence or absence of a band indicating hydrated minerals, rather than describing band shapes or doing detailed mineralogy. As measurements of asteroids in the 3-$\mu$m spectral region have been collected, it was recognized that a variety of band shapes were present in the population: Rivkin et al. (2003) found 375 Ursula to have a band shape interpreted as similar to Ceres, and 24 Themis was interpreted to have water ice frost and organic materials based on spectral modeling of its 3-$\mu$m band (Rivkin and Emery 2010, Campins et al. 2010). Informal taxonomies have been created to describe the spectral shapes seen in the literature. Takir and Emery (2012) identified 4 groups, while Rivkin (2010) favored a smaller number. In both taxonomies, however, a Ceres-like group and a Themis-like group is identified. Figure 1 shows the different band shapes we use in this work: "Ceres type", "Themis type", and "Pallas type", with the latter represented in Figure 1 by both Pallas itself and the asteroid 51 Nemausa. Interestingly, in a survey of 33 B-class asteroids, Clark et al. (2010) found Themis and Pallas to also represent major groups in that taxon based on their 0.5-2.5 $\mu$m spectra. None of the B or Cb asteroids larger than 200 km are Pallas types in the 3-$\mu$m region, save for Pallas itself.

The composition of Ceres based on the shape of its point-source 3-$\mu$m band shape (and the ~3.1-$\mu$m band in particular) has gone through several interpretations. Lebofsky et al. (1981) first suggested that the 3.1-$\mu$m band was due to a thin ice frost, but this interpretation lost favor due to thermodynamic considerations. The interpretations that followed included ammoniated phyllosilicates (King et al. 1992), carbonates and cronstedtite (Rivkin et al. 2006), and brucite and carbonates (Milliken and Rivkin 2009). The most recent interpretation of Ceres' spectrum, based on data from the Dawn mission, includes a mix of ammoniated phyllosilicates, carbonates, a dark component, and clay minerals (De Sanctis et al. 2018). It has been suggested that the presence of ammoniated minerals is an indicator that Ceres accreted material formed in the outer solar system or was formed there in its entirety (McKinnon 2012, De Sanctis et al. 2015).

As noted, Rivkin et al. (2003) found the asteroid 375 Ursula to have a 3-$\mu$m spectrum that was similar to that of Ceres, the first identified non-Ceres instance of such a band shape. While this band shape would now be characterized as "Themis-like" or "rounded" in the informal taxonomies, it provided the first hint that Ceres was not unique in its hydrated minerals. Takir and Emery (2012) presented 3-$\mu$m spectra of 28 asteroids, classifying 10 Hygiea and 324 Bamberga as "Ceres-like". In Section 3, we report new, additional spectra of Hygiea and Bamberga, as well as spectra from other large (D>200 km) objects with band minima between 3-3.15 $\mu$m, taken as part of a long-term spectroscopic survey. The Ch-class asteroids, which have 3-$\mu$m spectral shapes like Pallas, were discussed in Rivkin et al. (2015b). Other objects with band centers longward of ~3.15 $\mu$m or shortward of 3.0 $\mu$m will be the focus of other works, as will objects outside of the C complex and objects with diameters smaller than 200 km, as we focus here on the largest objects that appear to be most compositionally similar to Ceres or Themis. We note that Themis itself falls just short of the 200 km threshold, as its diameter according to the JPL Small-Body Database Browser is 198 km.

## 3 Observations

The data presented here were collected in the L-band Main-belt and NEO Observing Program (LMNOP: Rivkin et al. 2014, Rivkin et al. 2015). Observations of some of these objects have also been published by Rivkin et al. (2003), and other earlier workers (Jones et al. 1990, Lebofsky et al. 1990).

LMNOP data are all obtained using the SpeX instrument (Rayner et al. 2003) at the NASA Infrared Telescope Facility (IRTF) in its long-wave cross-dispersed ("LXD") mode in the short wavelength (1.9-4.1 $\mu$m) setting. A 15" slit is used, with a beam switch of 7" between A-B pairs of images. The specific exposure time and number of co-adds varies depending on the specific observing conditions, but the time between beam switches is kept below 120 seconds in order to allow subtraction of A-B pairs to correct for the effects of minute-scale changes in atmospheric conditions. The limiting factor for exposure time of an image (or coadd) is typically thermal emission from the atmosphere. Several solar-type standard stars are observed in a typical night, with airmasses matched as closely as possible to the asteroid observations. However, as will be discussed in Section 4, part of the reduction pipeline minimizes the effect of airmass mismatches. SpeX was upgraded in 2014, but all of the data presented and discussed here was taken with the pre-upgrade SpeX.

Table 1 shows the observing circumstances for the objects discussed in following sections, including V magnitude, distances from the Sun and Earth, phase angle, and sub-Earth latitude and longitude where available. The JPL Horizons ephemeris system provided the sub-Earth coordinates for Ceres, the Database of Asteroid Models for Inversion Techniques website (DAMIT: http://astro.troja.mff.cuni.cz/projects/asteroids3D/web.php) provided sub-Earth coordinates for Hygiea, Euphrosyne, Europa, and Thisbe. The remaining asteroids (Bamberga, Patientia, and Interamnia) are not available in the DAMIT database. However, calculations of sub-Earth latitude are straightforward for an object given its pole position and ephemeris position on a given date. Calculations of longitudes require precise knowledge of the rotation rate, and we do not provide estimates for sub-Earth longitude for these three asteroids. Table 2 compiles additional data about all the objects.

Only two of the objects have collisional families, and only 88 Thisbe has a density above 2.2 g/cm³ (though Thisbe and several others have large uncertainties for their density values).

# 4 Reduction

Reduction of LMNOP data has several steps. Extraction of spectra was done with Spextool (Cushing et al. 2003), a set of IDL routines designed for SpeX reduction developed and provided by the IRTF. After extraction, every combination of asteroid spectrum and star spectrum is run through an IDL-based set of routines developed and provided by Bobby Bus and Eric Volquardsen of the IRTF that correct for sub-pixel shifts between asteroid and star observations. It also uses an ATRAN model of the atmosphere (Lord 1992) to estimate the amount of precipitable water at the time of observation for each asteroid and star combination and remove it. This process has been used in several projects using SpeX data in the 0.8-2.5 and 2-4 $\mu$m regions (Clark et al. 2004, Sunshine et al. 2004, Rivkin et al. 2006, Rivkin et al. 2015b, among others). Following this step, a weighted average of the spectrum for each corrected asteroid-star pair was created for each asteroid, leading to a final asteroid spectrum. Bad pixels are flagged and omitted from the averaging process.

The temperatures of main-belt asteroids are sufficiently high to show detectable thermal emission in LXD spectra, particularly at their long-wavelength end near 4 $\mu$m. As a result, a correction to remove thermal emission is also made, which has a side benefit of providing some information about target thermal properties. The correction is made using a version of the Standard Thermal Model (Lebofsky et al. 1986), modified to allow values of the beaming parameter ($\eta$) to vary rather than remain fixed. Other than the beaming parameter, the other inputs to the STM are either observational circumstances (solar and Earth distances, phase angle, etc.) or physical properties (albedo, diameter, phase coefficients, etc.) that are well known for these large asteroids. The thermal contribution to the spectrum is forward modeled for a range of $\eta$ and a value is chosen that provides long-wavelength continuum behavior that is consistent with expectations (for instance, continuum behavior like meteorite analogs if identified or with photometric measurements if available: see Rivkin et al.(2013) and Rivkin et al. (2018) for fuller discussions of continuum selection). In the 3-$\mu$m band region the thermal contribution to the total flux is generally small. It is also monotonically increasing, so spurious band centers will not be introduced into the data by incorrect thermal models. If thermal flux is incorrectly estimated, the result will be incorrect spectral slopes, with the largest effect at the longest wavelengths in the spectrum (for instance in trying to identify the carbonate bands near 3.8-3.9 $\mu$m). However, incorrect thermal flux removal should not change the classification of a band shape from one of the groups considered here to another one. Figure 2 shows the spectrum of Bamberga from 30 April 2007 and 29 September 2013 before and after thermal flux removal. The inputs to the STM and the chosen values of $\eta$ to fit the thermal flux correspond to sub-solar temperatures for Bamberga of 219 K and 317 K for the 2007 and 2013 dates, respectively. Figures 3-10 show the thermal-removed spectra of the targets, grouped by object and offset from one another for clarity.

Spectra of Bamberga from three nights show apparent features with minima near 3.6-3.7 $\mu$m. Because at least one asteroid spectrum from two of those three nights exists with a positive feature in the same location as the apparent absorption in Bamberga, we interpret the features on those nights as artifacts likely due to the switch from the first to second grating order near those wavelengths. The third night, 27 June 2012, does not have any other asteroids showing this feature either as an "absorption" or "emission", nor does it show up in ratio spectra of the standard stars. However, while an absorption near 3.64 $\mu$m would be intriguing, we conservatively assume that given the other nights it is more likely to be an artifact than not. Nevertheless, given the hemispheric-level variation that Bamberga exhibits (Section 5.5), additional observations of Bamberga to verify that these features are artifacts may be worthwhile. We note that this same type of probable artifact is also seen in the spectrum of Interamnia on 12 September 2007 giving rise to the peak near 3.52 $\mu$m, with one other object from that night exhibiting the same behavior. There are hints of the same probable artifact in some other spectra, though they are not as prominent as the ones discussed here. The nature of this artifact is not fully understood, but we suspect it is related to the airmass at which an object is observed. It is only rarely seen in the dataset at a level comparable to the size of observational uncertainties, and has not been seen near 2.96 $\mu$m and near 2.54 $\mu$m at other grating crossovers. We leave these suspected artifacts in the figures for completeness and transparency, and note that the affected wavelengths are far from those we are most concerned with in this work.

# 5 Analysis and Results

As noted in Section 2.2, several different band shapes are seen in the 3-$\mu$m region on low-albedo asteroids. In the absence of a formal taxonomy, we analyze band shapes in three ways: first by visual inspection and sorting into groups, second by band centers and depths calculated via polynomial and Gaussian fits. Finally, we compare the band depths of the objects as measured at two different wavelengths in order to more quantitatively compare band shapes to the band shapes of type objects.

## 5.1 Visual inspection

The 3-$\mu$m spectral groups identified by Takir and Emery (2012) and Rivkin et al. (2012) are similar but not identical. The objects shown in Figure 1 represent the three 3-$\mu$m classes we use in this work, described as follows: We define Ceres-like objects here as having a relatively sharp local minimum near 3.07 $\mu$m and additional absorptions both longward and shortward of that wavelength (with an implied band minimum in the atmospheric opaque region, 2.5-2.85 µm) as contrasted with Pallas-like spectra, which have a monotonically-increasing reflectance from the long-wavelength end of the atmospheric opaque region until the continuum level is reached (usually near 3.2-3.3 $\mu$m), and Themis-like spectra, which based on the model fits from Rivkin and Emery (2010) and Campins et al. (2010) have a broad minimum near 3.1 $\mu$m or longward, while any additional band minima in the atmospheric opaque region are either absent or sufficiently narrow to be unobservable in ground-based spectra. A further possible discriminator between Ceres- and Themis-types is an absorption near 3.8-3.9 $\mu$m due to carbonates seen in Ceres' spectrum, but data quality is not always high enough to detect this band, and small changes in thermal flux removal could affect its interpretation. Therefore, in this work we do not consider its absence diagnostic.

Using these criteria, we identify the qualitative taxonomic assignments shown in Table 3.

To our knowledge, spectra of hydrated meteorites in the literature have Pallas-type band shapes as a rule (Rivkin et al. 2015a), particularly in the wavelength region accessible to ground-based observations. In a qualitative sense, Pallas-type band shapes are thought to be due to phyllosilicates, which have band minima in the 2.7-2.8 $\mu$m region blocked from the ground by atmospheric water vapor. It is presumed that other Ceres-type objects share the surface composition of Ceres (or a similar composition) in the absence of other information to the contrary. The composition of the Themis types is interpreted to include water-ice frost and organics (Rivkin and Emery 2010, Campins et al. 2010), though the expected short lifetime for ice on asteroidal surfaces has led to alternate proposed compositions such as goethite (Beck et al. 2011) or ammoniated minerals (Brown 2016).

5.2  Quantitative Band Measurements:

Although visual inspection has been used in the past and in the previous section, detailed analysis requires more quantitative study. While Pallas-type spectra can be easily distinguished from other spectra based on reflectance ratios, it is much more difficult to distinguish Ceres- and Themis-types from one another on this basis, especially given typical uncertainties. Principal component analyses (Bus and Binzel 2002) and neural network analyses (Howell et al. 1994) have been used in the past on asteroid spectra and may be of use in creating a taxonomy for 3-$\mu$m spectra, but are beyond the scope of this paper. Instead, we look to simpler tools to describe the data discussed here. We divide the 3-$\mu$m region into three areas: "Region 1" shortward of 3.0 $\mu$m, "Region 2" from 3.0-3.25 $\mu$m, and "Region 3" between 3.25 and 3.5 $\mu$m.

*Polynomial Fitting:*

We used a 6$^{th}$-order polynomial fit to the 2.9-3.4 $\mu$m region for each of the observations of C-complex asteroids in the LMNOP. This was done strictly as a means of estimating the position of the band minimum and band depth for the deepest (and in some cases second-deepest) absorption. These quantities can be estimated even for poor quality data, but obviously are less secure for such observations. This approach will only identify broad features, but sharp features are not typically seen in asteroid spectra. All LMNOP objects with any spectra with a band in Region 2 and with a diameter > 200 km are discussed here.

Table 3 shows the band depths and centers found between 2.9-3.4 $\mu$m as calculated from the best-fit polynomials, and Figure 11 shows the polynomial fits to Ceres, Themis, and Pallas. These were found by calculating the polynomial value every 0.005 $\mu$m, and thus they have an uncertainty of ±0.0025 $\mu$m, outside of any additional uncertainties due to the fitting technique (including departure of the true reflectance spectrum from polynomial behavior) itself. Although this choice of polynomial sampling frequency is somewhat arbitrary, it is a good match to the original spectral resolution provided by SpeX in LXD Short mode (R~937.5 with the 0.8" slit). In some cases, no band was identified in Region 3. For two spectra of 704 Interamnia, an absorption continued to deepen beyond 3.4 $\mu$m, where the polynomial was no longer fitted. In these cases, the band center in Region 3 is given as 3.40 $\mu$m in Table 3 and the band depth at 3.40 $\mu$m is provided.

*Gaussian fitting:*

In addition to the polynomial fits, absorptions in Region 1 and Region 2 were fit for spectral slope-removed, normalized spectra with a constant-reflectance continuum and a single Gaussian in order to provide a characterization independent of the polynomial fit. The spectral slope removal step used fixed wavelengths as endpoints, and rather than have the spectral slope removal affect the best fit for Gaussian width and amplitude, the reflectance of the constant-reflectance continuum (i.e. the baseline from which the Gaussian was measured) was treated as a free parameter. In all cases the continuum value remained within 5% of 1, and in 30 of 34 cases it is within 2.5% of 1. The data points used to fit the Gaussian were typically restricted to some points shortward of 2.5 $\mu$m, and data between 2.9 and 3.25 $\mu$m. The bands seen in the 3.3-3.4 $\mu$m region are not well-described with a single Gaussian, and were not fit using this technique. In some cases where there was no identifiable band in Region 3, a longer-wavelength cutoff was used for the data used to fit absorptions in Region 2. Table 3 also shows the center and the amplitude of the Gaussian fits for Region 2, taken to represent the band center and band depth, along with the fitted uncertainty for wavelength of the Region 2 Gaussian. In most cases the specific range of data points used for the Gaussian fits had negligible influence on the output band parameters, but the spectrum of 451 Patientia from 30 October 2010 had significantly different results depending upon the range of data points used, and as a result we do not report a fit. Because the bands are absorptions, the amplitude of the Gaussians is a negative number. However, we report the absolute value of the amplitude to be consistent with the general conception of a band depth as a positive number. In three cases (Hygiea on 18 May 2005 and Interamnia on 17 September 2006 and 29 August 2012) the strongest absorption, and the one reported in Table 3, was in Region 1 instead of Region 2.

*Results:*

The band parameters found for the Gaussian and polynomial fits in Table 3 are in generally good agreement (Fig. 12). In most cases they agree within ~0.02 $\mu$m for band center and 0.02 for band depth. The cases where they do not agree largely correspond to situations where the band center is near or shortward of 2.95 $\mu$m, or where the data quality is among the worst in the overall dataset. The correlation coefficient between the two approaches to measuring band centers is 0.80. Figure 13 shows a representative case, a spectrum of Euphrosyne with its polynomial and Gaussian fits overplotted.

As expected, the spectra identified as Pallas types have reflectance minima at or near 2.9 $\mu$m, the short end of the wavelength range considered. This does not necessarily indicate a band center at that wavelength, however, as Figure 1 shows that in these cases reflectances continue to decrease into the atmospheric opaque region. In these cases, Table 3 shows the band center as "<2.95 $\mu$m" and the band depth as greater than the 2.95-$\mu$m band depth for the polynomial-based fits.

There are two striking results. First, the band centers found on a single object can vary. Interamnia has spectra with band centers in both Region 1 and Region 2. Bamberga has spectra with absorption band centers ranging from 3.04 $\mu$m to 3.20 $\mu$m in the Gaussian fits. Hygiea's band centers are largely near 3.05-3.08 $\mu$m, but include band centers at shorter wavelengths. This variation on single objects is discussed in more detail in Section 5.5.

The second result is quantifying the variation exists from one object to the next. Figure 14 is a histogram of Region 2 band centers for both the polynomial and Gaussian fits. While there appears to be a bimodal distribution in band centers, there is no obvious gap over the 3.03-3.12 $\mu$m range that can be used to separate the Themis-types from the Ceres-types. We note the most common band center seen in the polynomial fits is the 3.04-3.05 $\mu$m bin, which includes the spectrum of Ceres itself. Interestingly, the bin containing Ceres is only the 4$^{th}$ most common band center in the Gaussian fits. Given the typical uncertainty on the Gaussian fits (Table 3), many of the band centers are within 1-$\sigma$ of each neighboring bin. Therefore, we are unwilling to ascribe much meaning to the relatively empty bin for Gaussian band centers from 3.07-3.08 $\mu$m, which abut the most-populated and 4$^{th}$-most-populated bins. More data, with smaller uncertainties, will be necessary to determine whether the distribution of band centers over the 3.03-3.12 $\mu$m range has distinct peaks or is a currently poorly-sampled single distribution.

The calculated band centers > 3.12 $\mu$m belong to Bamberga, Thisbe, Europa and Euphrosyne in the polynomial fits, with only Bamberga and Europa having Gaussian fits at those wavelengths, though the Thisbe and Euphrosyne band center positions overlap for the two approaches within uncertainties. Figure 13 shows the spectrum of Euphrosyne from 21 September 2005 with the polynomial and Gaussian fits overlaid, demonstrating the sensitivity of the polynomial to asymmetries in the band shape, though it is not clear whether the band in Region 3 traced by the polynomial fit but missed by the Gaussian fit is real or noise.

The Gaussian fits also produce a value for the half-width at half-maximum for absorptions in each spectrum. These range from 0.1 or less for the sharpest bands to 0.35 for the 17 May 2006 spectrum of Europa. Like band centers, there is no obvious value at which the Ceres- and Themis-types can be distinguished. Furthermore, it is not obvious how best to represent the presence/absence of an absorption band centered in Region 1, which could discriminate between Themis- and Ceres-types, because the band centers have a range of wavelengths and the width of the atmospheric opaque region can be variable with the quality of the observing night. In short, it is not obvious from these data how to separate Themis-types from Ceres-types on the basis of simple numerical parameters. More sophisticated techniques, like principal component analysis or machine learning approaches, may generate useful taxonomies but are beyond the scope of this work.

Looking beyond the asteroid population, there is not only an apparent continuum between Themis- and Ceres-types in terms of band center, but also potentially between Themis-types and comets. Figure 15 shows the continuum-removed spectra of several objects in the sample as well as Comet 67P/Churyumov–Gerasimenko from Rosseau et al. (2018), showing the variation in band centers across these objects as well as the qualitative similarity of Bamberga's 20 March 2002 spectrum to that of 67P (see also the discussion of Bamberga in Section 5.5). Additional work will be necessary to determine the Bamberga's composition, and whether this qualitative similarity is meaningful.

## 5.3 Compositional Interpretation

As noted, while there are few gaps in the distribution of band centers, there are clusters near certain wavelengths. Two Interamnia spectra and one Hygiea spectrum have their deepest absorption near or shortward of 2.95 $\mu$m, in Region 1. These are likely due to phyllosilicates, as seen in C chondrite spectra and the Pallas-type spectra discussed above (Sato et al. 1997, Rivkin et al. 2015b). Next is a cluster of band centers in the 3.03—3.09 $\mu$m range within Region 2 with peaks in the distribution near 3.05 $\mu$m and 3.08 $\mu$m in the polynomial fits and 3.06 $\mu$m and 3.09 $\mu$m in the Gaussian fits. The first of these peaks includes the spectrum of Ceres from 18 September 2006, and is close to the average band center for Dawn VIR data reported by Ammanito et al. (2016) (3.061 $\mu$m ± 0.011 $\mu$m). These absorptions are most straightforwardly interpreted as due to ammoniated minerals, like those of Ceres are interpreted by De Sanctis et al. (2016). Ehlmann et al. (2018) report absorptions throughout this range in spectra of smectites that were ammoniated in the laboratory, attributing the variation in wavelength to the degree of hydrogen bonding of $NH_4$-$H_2O$ complexes.

An additional peak in the distribution is found within Region 2 beyond 3.12 $\mu$m. These objects are spectrally similar to Themis within observational uncertainties, consistent with them having similar composition to what is interpreted for Themis—ice frost and organic materials on a low-albedo, anhydrous silicate surface, though it is possible that hydroxyl absorptions may be present in the spectral region obscured by the atmosphere. Themis itself has a band center slightly shorter than these wavelengths, however, as seen in Table 3. It is not obvious whether these absorptions are at wavelengths too long to be due to ice frost. The absorption band seen at 3.20 $\mu$m in the March 2002 spectrum of Bamberga is likely due to some other source. The broad absorption on 67P centered near 3.25 $\mu$m is not attributed to a single species, with Quirico et al. (2016) finding aromatic C-H, COOH groups, OH groups, and $NH_4^+$ as plausible carriers of the band and favoring the carboxylic group as most plausible. These may also be present and responsible for some of the features in Themis-type spectra in addition to or rather than ice frost.

Absorptions in Region 3 near 3.3-3.4 $\mu$m are seen in many (but not all) of the spectra included here. Carbonates and organic materials both have absorptions at these wavelengths, and they can be difficult to separate from one another. These absorptions were interpreted in the spectrum of Themis by Rivkin and Emery (2010) and Campins et al. (2010) as due to organic materials. Over most of Ceres' surface, absorptions at those wavelengths are assigned to carbonates (Rivkin et al. 2006, Carrozzo et al. 2018) because they are accompanied by a 3.9-μm carbonate band, which is not seen in the Themis spectrum. In localized patches found by Dawn, the bands are augmented by organic material as well (De Sanctis et al. 2017, Pieters et al. 2018, Kaplan et al. 2018).

The final non-spurious features seen in the asteroids discussed here are features beyond Region 3, near 3.8-3.9 $\mu$m and also attributed to carbonates in the spectrum of Ceres. As noted, we do not attempt to quantitatively measure these features in this work but there are hints of their presence in some of the highest-quality spectra, including the 13 September 2007 and 27 June 2012 spectra of Hygiea.

Also as noted, there are some apparent features near 3.5-3.6 $\mu$m in several spectra, but we interpret these as artifacts rather than having compositional information.

Finally, we note that the effects of space weathering on these objects are largely unknown. Lantz et al. (2017) found that the phyllosilicate bands in carbonaceous chondrites shifted to longer wavelengths after ion irradiation, but it is not certain whether such shifts would be expected in the materials present on the asteroids we are discussing. If their spectra are affected in the way that silicate spectra at shorter wavelengths are affected, we might expect a decrease in band depths after exposure, regardless of wavelength region. It is conceivable that some of the band depth differences seen from object to object or on a single object could be evidence of differences in time of exposure to space weathering processes rather than related to the amount of absorbing material. However, we would not expect space weathering to cause a Pallas-type band to transform into a Ceres- or Themis-type band or vice versa: the absorptions that give these bands their shapes can be attributed to particular minerals, as discussed throughout this section, and space weathering processes are not thought to create these minerals.

5.4  Band Depths and Band Shapes

Moving beyond band centers and depths, we also wish to characterize band shapes. A relatively simple approach is to look at the band depths found at particular wavelengths that are not necessarily near band centers. We follow the example of earlier studies of carbonaceous chondrites and C-class asteroids (Sato et al. 1997, Rivkin et al. 2003, Rivkin et al. 2015b), and look at band depths at 2.9 and 3.2 $\mu$m.

Figure 16 shows the 2.9-$\mu$m band depth vs. 3.2-$\mu$m band depth for the spectra in Table 3, along with spectra of Ceres, Themis, and 51 Nemausa (from Fig. 1), representing endmembers of 3-$\mu$m band shapes we have seen thus far (as discussed in Section 2.2). Also included are Comet 67P and Pallas itself, along with a set of Ch-class asteroids from Rivkin et al. (2015b) and CM meteorite spectra from Takir et al. (2013). Finally, lines are drawn from each of the three endmember points and from 67P to the origin, showing roughly where mixes of each endmember with a neutral material would fall. Nemausa is used as the Pallas-type endmember here rather than Pallas because of Nemausa's particularly deep 3-$\mu$m absorption band.

As was shown in Rivkin et al. (2015), the Ch meteorites and CM meteorites fall in the same region of this plot, with 2.9-$\mu$m band depths greater than 3.2-$\mu$m band depths by a factor of roughly 2-3. Nemausa, itself a Ch asteroid, is at the far end of this region, but is accompanied by the Takir et al. spectrum of meteorite LAP 03786. This region includes Pallas, as well. Several of the objects, including one of the spectra of Interamnia, are within 1-$\sigma$ of the Nemausa-neutral mixing line. At the other extreme is 67P, which has a relatively large band depth at 3.2 $\mu$m and very little at 2.9 $\mu$m. Bamberga has spectra that are consistent with 67P, or are even further from the origin in its direction.

While several spectra fall in the vicinity of Themis, the area around Ceres is relatively empty, save for one of the Hygiea spectra. Indeed, on this particular figure, the 2[nd] and 3[rd]-closest spectra to Ceres belong to a CM meteorite and a Ch asteroid. The spectra of Euphrosyne fall on or near the Themis-neutral mixing line, while two spectra of Europa appear to fall near the extension of that line beyond Themis. In a broad sense, we can say that Interamnia's spectra tend to fall with the Pallas types or between the Pallas types and Ceres, some of Bamberga's spectra have similarities to 67P (as noted), and the remainder appear consistent with Themis or are intermediate between Themis and Ceres. The majority of the spectra from Table 3 fall in a region bounded by the line segments from the origin to Nemausa and from the origin to Themis and beyond, suggesting that, at least in a simplistic sense, it may be possible to mix the spectra of Nemausa, Themis, and Ceres and a neutral component to recreate the spectra in-between. However, the presence of Bamberga spectra beyond those bounds suggests that those four end members are insufficient to recreate all spectra in the sample.

5.5 Variation on objects

Reports of spectral variation on asteroids have long been controversial. Studies of Ceres from ground-based, space-based, and Dawn measurements show that body to have little spectral variation on very large spatial scales (Carry et al. 2012, Schäfer et al. 2018), though Dawn has found some smaller-scale variation. Dawn measurements at Vesta showed deeper bands in the 3-$\mu$m region associated with large-scale lower-albedo areas (De Sanctis et al. 2012), but those variations would have been challenging to conclusively detect in the unresolved spectra measurable from the ground.

Given these results, the large-scale differences seen in the spectra of some objects are unexpected. We note that such differences are not seen in every object: for instance, 31 Euphrosyne and 52 Europa show no variation within observational uncertainties despite four observations per object. Uncertainties in pole positions and rotation periods can make it difficult to extrapolate the central latitude/longitude from apparition to apparition, but here we call out three cases of apparent surface variation in the hydrated mineralogy of large low-albedo asteroids: 10 Hygiea, 704 Interamnia, and 324 Bamberga.

*10 Hygiea:* Hygiea has been identified as having a Ceres-like spectrum (Takir and Emery 2012, Rivkin et al. 2014). A cursory inspection of Figures 3 and 16, however, makes it clear that it does not always present a spectrum like Ceres. Figure 16 shows some spectra are more similar to the Ceres endmember and others to the Themis endmember, though as noted that is not necessarily indicative of compositional differences. Tables 1 and 3 show the sub-Earth coordinates for Hygiea for the available observations (which also appear on Figure 3) as well as the band centers and band depths.

Sub-Earth and sub-solar coordinates for the Hygiea observations are available from the DAMIT website (Durech et al. 2010). There is no correlation seen between Gaussian-fit band center and longitude, though there is a moderate correlation for longer-wavelength bands to be found at more positive latitudes. This correlation seems to be driven by the 18 May 2005 observation, however, and the polynomial fit finds a very different band center for that date (Figure 17). Including the measurements of Takir and Emery also removes the correlation for the Gaussian fits. Additional high-quality measurements are needed to determine whether the departures from Ceres-like behavior are real or due to noisy data.

*704 Interamnia:* Surprisingly, given its size, Interamnia has no shape model in the DAMIT database, and as noted we consider only latitudinal variation for it, using the Drummond et al. (2008) pole position (Figure 18). The spectrum from 12 September 2007 is the one that deviates most from the general appearance of the other spectra, though the other spectra also show some variation from more Pallas-type to those having some Ceres-type features. This is borne out by the polynomial and Gaussian fits, which show features at 3.05 ± 0.03 $\mu$m in the 1 September 2012 and 11 December 2013 spectra.

However, there is some reason to doubt at least some of this variation. The 29 August 2012 and 1 September 2012 observations took place a little under 3 days apart from one another, close enough in time to calculate their relative rotational phases with good accuracy and find they were observed ~0.09 of a rotation from one another. In addition, their sub-Earth latitudes were rather polar (-69 degrees), which minimizes the ability to detect rotational variability. As seen in Figure 10, the differences between the two spectra are relatively subtle, which likely plays a role in the different band center measurements. Further observations will be necessary to establish whether variation is observable at that sub-Earth latitude.

As noted, the data from 12 September 2007 shows the largest difference from the others, and appears to have a rather Ceres-like spectrum with a minimum near 3.085 $\mu$m and what visually appears to be a second band near 3.35 $\mu$m, though the order-crossing artifact near 3.52 $\mu$m makes that interpretation more difficult. These data were taken when Interamnia was at a somewhat more equatorial aspect than the 2012 data, though the rest of the data are at still more equatorial aspects, suggesting that the region causing the very Ceres-like spectrum may be confined in longitude rather than a latitudinal effect. A Ceres-like spectrum was also seen in the 2013 observations, but the elapsed time between the 2007 and 2013 observations is too long to determine whether similar longitudes were observed. A dedicated observing campaign to observe Interamnia over a full rotation at a time when it is near an equatorial aspect will likely be necessary to confirm if the spectral shape seen on 12 September 2007 is due to longitudinal variation.

*324 Bamberga:* The situation for Bamberga appears to be more straightforward than either Hygiea or Interamnia. Table 1 shows the sub-Earth latitude for Bamberga for the available observations, the band centers and depths are on Table 3, and Figures 7-8 show the associated spectra, with Figures 19-20 showing the band center vs. sub-Earth latitude and band depths vs. sub-Earth latitude. It can be seen by inspection that observations of Bamberga's southern hemisphere have deep features, including absorptions at relatively longer wavelengths than most objects in the sample. By comparison, observations of Bamberga's northern hemisphere have a much smoother spectrum in the 3-$\mu$m region, with a much-reduced absorption depth. This pattern holds over decades, and includes observations by Takir and Emery (2012) and Rivkin et al. (2003), the latter of which utilized data from a different instrument and telescope than the remainder of measurements.

We can reject two possible causes for the difference: thermal fill-in and incorrect thermal flux removal. While the observations in 2013 were made at solar distances < 2 AU, the thermal flux was only about 6 times the reflected flux at 3.6-3.7 $\mu$m and much less at shorter wavelengths, far short of the ~20 times that would be necessary to fill in the absorption bands seen in the 2007 and 2012 measurements (Rivkin et al. 2013). Conversely, the nature of the thermal modeling and correction is such that absorption bands cannot be created by poor thermal correction—if the models over-correct, the over-correction increases with increasing wavelength rather than reversing to create an apparent band.

The most straightforward interpretation of the variation in these spectra is that there is hemispheric-scale spectral variation with latitude on Bamberga. No shape model for Bamberga appears in the DAMIT database, so no longitude information is available, but the dataset does not appear to require any longitudinal variation. Given our limited understanding both of what compositions are responsible for the bands we see on Bamberga and how their spectra might change from space weathering processes, it is unclear what the cause of the variation might be, but it is at least qualitatively consistent with what we might expect from differences due to composition and exposure/age.

Interestingly, and as noted, while the spectra from the southern hemisphere of Bamberga have absorptions with band centers at wavelengths longer than what is seen in Ceres or Themis, as Figure 15 shows they make an intriguing comparison to spectra of comet 67P (Rousseau et al. 2018). We note for future work that cometary non-ice surface compositions may be quite similar to large low-albedo asteroid compositions.

## 5.6  Further Implications of Variability

While we recommend additional measurements of Hygiea and Interamnia to characterize any surface variation, it appears that both objects have areas with Pallas-type and non-Pallas-type spectra. Models of Ceres' interior suggest that there may be a hydrated silicate layer beginning at a depth of ~50 km below a low-density clay/clathrate layer (King et al. 2018, Castillo-Rogez et al. 2018). If Hygiea and Interamnia also have compositional layers, it is possible that Hygiea had its interior exposed in only a few places, leading to a mostly Ceres-type spectrum with a few Pallas-type areas, while Interamnia had much of its original surface removed save for a few small regions. On the other hand, it is not clear whether the interior of Ceres (or of Ceres-like objects) would have a Pallas-type spectrum if exposed. Further geochemical work will be needed to determine if this could explain the co-incidence of Pallas-type and Ceres- or Themis-type spectra on the same object.

Some objects also have both Ceres- and Themis-like spectra. This in some cases, like for 451 Patientia, may be a matter of data quality. Taken at face value, we can look at the presence of ice at the surface of Ceres in localized areas (Combe et al. 2016) and its high concentration in the near subsurface (Prettyman et al. 2017). If the objects in our sample are similar, we can imagine they might have large areas where ice is present (perhaps temporarily?) with a near-surface reservoir that could replenish it under some conditions. Alternately, if the suggestions that Themis-like bands are not due to ice are borne out (see further discussion in Section 5.7), Themis- and Ceres-type bands could plausibly represent compositional differences, perhaps different amounts of ammoniated or hydrated minerals and/or different amounts of space weathering.

## 5.7  Ceres- and Themis-types: Meteorites and Collisional Families

The asteroids we consider here with Ceres- and Themis-like spectra have spectral properties that are not represented in the meteorite collection. Mg-Fe phyllosilicates dominate the hydrated mineralogy of the aqueously altered carbonaceous chondrite groups. Even if we consider the identifications of the compositions of Ceres and Themis to be unsettled, it is clear that there are important differences between their 3-$\mu$m spectra and the CM meteorites and Ch asteroids, and those differences also can be seen among the eight large C-complex asteroids we discuss and the meteorites. Objects are delivered to the near-Earth object population from the entire asteroid belt, albeit with different efficiencies for different parts of the belt (Granvik et al. 2017). Given the relative abundance of these objects, it is not obvious why we should lack meteorites that are consistent with the Ceres- and Themis-types but have them from not only Pallas-types but also, apparently, the high-albedo objects > 200 km (Vesta and HEDs, and several S-class asteroids and ordinary chondrites, mesosiderites, etc.).

There seem to be three broad reasons why these objects, which make up 36% of the mass of the main asteroid belt (8.5% if Ceres is removed from consideration) might not contribute to the meteorite collection: 1) Possible meteorites cannot penetrate the atmosphere; 2) Pieces of these objects do not make it to near-Earth space, 3) They are present in the meteorite collection, but we cannot recognize them.

Further study of these objects, and smaller members of the main belt, would likely distinguish between these possibilities. The second possibility would be technically eliminated if a near-Earth object with Ceres- and/or Themis-like spectrum were found, and the first would be favored if such objects were found in numbers roughly consistent with expectations from delivery calculations. The third possibility would be harder to test per se, but would require either the composition responsible for the absorptions to be relatively rare volumetrically on the parent body and thus unlikely to be found in meteorites, unstable at 1 AU and thus absent by the time of Earth impact, or rapidly altered after arrival at Earth (like the halites seen in Zag and Monahans: Zolensky et al. 2000) and thus absent from meteorites unless collected in conditions designed to preserve them. Of course, multiple factors could be relevant. Water ice, for instance, is unlikely to be preserved in any but the largest NEOs (possibility 2, Schorghofer and Hsieh 2018), and even if present its preservation would be very unlikely after arrival on the Earth (possibility 3).

Rivkin et al. (2014) demonstrated that Ceres would very likely have suffered impacts of similar size to those that made collisional families for Vesta, Pallas, and other objects, and argued sublimation of ice-rich ejecta after large collisions could explain Ceres' lack of a dynamical family. It is tempting to apply the same logic to the objects in this sample. Most of the objects discussed here also lack families (Table 2), though it is possible they happened to escape a large enough collision to create one. Hygiea and Euphrosyne both are associated with dynamical families, as is Themis. Measurements of the Themis-family member 90 Antiope shows a similar 3-$\mu$m band shape as Themis itself (Hargrove et al. 2015), but measurements of smaller members of the Hygiea and Euphrosyne families are beyond the capabilities of current groundbased facilities and instruments, which are challenged to measure the 3-$\mu$m spectra of objects fainter than V ~ 14-15. The Themis family formed via a catastrophic impact, and the reaccumulated pieces of an ice-rock differentiated object may have inherited different amounts of ice (Castillo-Rogez and Schmidt 2010).

Hygiea appears near the edge of its dynamical family, which can be interpreted as a sign of a very oblique impact or a hint that Hygiea is itself an interloper (Carruba 2013, Rivkin et al. 2014). If the smaller members of the Hygiea family are observed to have 3-$\mu$m spectra similar to Hygiea, it would favor Hygiea's identification as the family parent body and also demonstrate that the minerals responsible for Hygiea's spectrum are not destroyed in family-forming impacts. If the variations seen in Hygiea's spectrum can be confirmed and traced to a particular location on its surface it may allow a connection to be made to the family-forming impact perhaps via improved imaging capabilities coming online with the James Webb Space Telescope (Rivkin et al. 2016) or adaptive optics on large telescopes like VLT/SPHERE (Marsset et al. 2017). Such work still is in the future, however.

Given the uncertainties in the Hygiea family, study of the Euphrosyne family may offer the most straightforward path. Masiero et al. (2015) identified a pathway to near-Earth space for the Euphrosyne family and listed a set of NEOs potentially derived from that family. If the 3-$\mu$m band on Euphrosyne is due to ice frost, we might expect its absence in smaller family members, particularly those that become NEOs and reach high temperatures. If measurements of candidate Euphrosyne-family NEOs show 3-$\mu$m band shapes similar to Euphrosyne itself, we could straightforwardly rule out option #2 above. It would also point away from ice frost being the cause of the Themis-like bands, given the very high temperatures experienced by NEOs compared to ice stability temperatures.

# 6 Summary and Conclusions:

Eight of the twelve C-complex asteroids larger than 200 km have been observed to have Ceres- or Themis-like spectra at least once. Polynomial and Gaussian fits to their spectra show a bimodal distribution in band centers, with peaks near the band centers of Ceres and Themis themselves. An absorption centered near 3.3-3.4 $\mu$m is also present in many of the spectra, though it is not seen on every object or in every spectrum of some objects. While most of the spectra have band shapes/centers that are consistent with Themis, Ceres, the Ch-asteroid Nemausa, or mixtures of them, Bamberga has some spectra that appear more similar to Comet 67P. Variation in spectral properties is seen on several of these objects, particularly on Bamberga, which shows consistent evidence of a northern-southern hemisphere spectral dichotomy.

The composition found at Ceres' surface is thought to involve significant aqueous alteration and in some areas, transport to the surface or exhumation after formation in Ceres' interior (Castillo-Rogez et al. 2018). The spectral similarity of some of these objects to Ceres opens the possibility that they also had similar histories, up to and including transport of minerals to their surfaces after subsurface aqueous alteration. The spectral similarity of others to Themis provides additional constraints for understanding the distribution of ice on asteroidal surfaces. The variation seen from object to object and from observation to observation on some objects allows the opportunity for further study of aqueous alteration, volatile minerals, and their evolution on asteroids and their interrelationship. Because these objects have very likely never suffered disruption, as noted at the outset, the variation helps constrain the spatial scale of aqueous alteration and/or dehydration processes that have occurred through their histories. The lack of meteorites that are apparently derived from these objects has implications for family formation and evolution, NEO delivery, and potentially even terrestrial weathering. Future observations from ground- and space-based observatories and spacecraft missions should be able to test the similarities and differences between these objects and the samples delivered by impacts and by spacecraft to determine if those similarities and differences are more than merely skin deep.

***Acknowledgements and Data:***
ASR acknowledges long-standing support from the NASA Planetary Astronomy program over the timeframe that these observations and analyses have been done, including grants

NNX14AJ39G, NNX09AB45G, NNG05GR60G, and NAG5-10604. This work, and companion papers past and future, would not be possible without such steady support. Grant SOF 04-0050 from SOFIA has also provided partial support for this work for ASR. JPE acknowledges SSO grant NNX16AE91G and generous support of the L.A. Taylor Endowment. The raw telescopic data that were used in this paper are held at the Caltech/IPAC Infrared Science Archive, and publicly available as of June 2019. The reduced spectra used in the figures are available at the Johns Hopkins Applied Physics Laboratory Data Archive at http://lib.jhuapl.edu/papers/infrared-spectroscopy-of-large-low-albedo-asteroid/ and are being submitted to the Planetary Data System Small Bodies Node.

All authors are Visiting Astronomers at the Infrared Telescope Facility, which is operated by the University of Hawaii under contract NNH14CK55B with the National Aeronautics and Space Administration. We acknowledge the sacred nature of Maunakea to many Hawaiians, and our status as guests who have been privileged to work there. Many thanks to the stalwart telescope operators of the IRTF who were instrumental in taking these data through the years, and to Bobby Bus and Eric Volquardsen for developing "the ATRAN part" of the data reduction. Thanks to Driss Takir for sharing his spectra with us for analysis. Thanks to Beth Clark and an anonymous reviewer who outed himself as Ralph Milliken for helpful reviews.

*Tables*

| Date | Asteroid | V mag | R (AU) | Δ (AU) | Phase angle | Sub-Earth latitude | Sub-Earth longitude | Exposure time (sec) | PW (mm) |
|---|---|---|---|---|---|---|---|---|---|
| 18 Sep 2006 | 1 Ceres | 8.20 | 2.98 | 2.17 | 13.2 | +2° | 347° | 720 | 4.05 |
| 23 Aug 2002 | 10 Hygiea | 11.53 | 3.47 | 3.17 | 16.7 | -11° | 241° | 960 | 3.70 |
| 18 May 2005 | | 10.28 | 2.83 | 2.16 | 17.5 | -26° | 183° | 960 | 1.50 |
| 8 Sep 2006 | | 10.52 | 2.95 | 2.32 | 17.3 | +45° | 261° | 1440 | 2.72 |
| 17 Sep 2006 | | 10.67 | 2.96 | 2.44 | 18.4 | +45° | 341° | 720 | 3.64 |
| 12 Sep 2007 | | 10.51 | 3.35 | 2.41 | 7.5 | +17° | 14° | 600 | 2.46 |
| 13 Sep 2007 | | 10.49 | 3.35 | 2.41 | 7.2 | +17° | 46° | 720 | 1.99 |
| 27 Jun 2012 | | 10.66 | 3.05 | 2.41 | 16.8 | +39° | 199° | 1800 | 0.80 |
| 21 Sep 2005 | 31 Euphrosyne | 11.24 | 3.03 | 2.07 | 6.8 | +15° | 132° | 1920 | 1.60 |
| 28 Aug 2011 | | 11.74 | 2.76 | 2.28 | 20.5 | -15° | 13° | 1200 | 1.33 |
| 30 Jan 2012 | | 11.67 | 2.51 | 2.33 | 23.1 | -29° | 73° | 1560 | 1.39 |
| 8 Jan 2013 | | 11.75 | 2.73 | 2.31 | 20.4 | -22° | 247° | 900 | 1.06 |
| 17 May 2006 | 52 Europa | 11.01 | 3.20 | 2.22 | 5.8 | -46° | 331° | 1280 | 3.06 |
| 28 Jun 2012 | | 11.28 | 3.38 | 2.40 | 5.5 | -58° | 165° | 1260 | 1.08 |
| 27 Jun 2013 | | 11.93 | 3.39 | 2.73 | 14.6 | -11° | 78° | 1170 | 4.44 |
| 21 Jul 2013 | | 11.52 | 3.38 | 2.48 | 9.4 | -12° | 70° | 1080 | 2.53 |
| 22 Jul 2009 | 88 Thisbe | 10.49 | 2.31 | 1.42 | 15.6 | +0° | 353° | 1536 | 3.47 |
| 7 Sep 2009 | | 10.09 | 2.33 | 1.35 | 7.6 | +2° | 111° | 540 | 5.60 |
| 20 Mar 2002 | 324 Bamberga | 11.93 | 3.39 | 2.45 | 6.1 | -31° | -- | 1200 | 1.23 |
| 30 Apr 2007 | | 12.40 | 3.58 | 2.70 | 8.9 | -39° | -- | 1440 | 0.91 |
| 27 Jun 2012 | | 12.34 | 3.20 | 2.58 | 16.0 | -34° | -- | 1710 | 0.72 |
| 3 Jul 2012 | | 12.42 | 3.19 | 2.64 | 16.9 | -34° | -- | 1170 | 0.99 |
| 27 Jun 2013 | | 10.45 | 1.99 | 1.42 | 29.0 | +25° | -- | 1260 | 3.48 |
| 21 Jul 2013 | | 9.79 | 1.92 | 1.14 | 25.5 | +28° | -- | 720 | 2.90 |
| 25 Aug 2013 | | 8.64 | 1.84 | 0.87 | 12.4 | +34° | -- | 720 | 0.99 |
| 29 Sep 2013 | | 8.44 | 1.79 | 0.82 | 11.7 | +40 | -- | 1260 | 1.87 |
| 31 Oct 2010 | 451 Patientia | 12.58 | 3.16 | 3.25 | 17.8 | +19° | -- | 1440 | 0.79 |
| 28 Aug 2011 | | 11.51 | 2.93 | 2.20 | 15.7 | -45° | -- | 1440 | 1.23 |
| 17 Sep 2006 | 704 Interamnia | 10.24 | 2.68 | 1.78 | 12.2 | -24° | -- | 1050 | 3.68 |
| 12 Sep 2007 | | 11.60 | 2.73 | 2.73 | 21.2 | -35° | -- | 600 | 2.05 |
| 29 Aug 2012 | | 11.13 | 2.61 | 2.25 | 22.5 | -69° | -- | 1710 | 2.07 |
| 1 Sep 2012 | | 11.09 | 2.61 | 2.22 | 22.3 | -68° | -- | 720 | 2.78 |
| 11 Dec 2013 | | 11.71 | 3.19 | 2.69 | 16.5 | +22° | -- | 600 | 3.28 |

**Table 1:** Observational Circumstances for the objects in this work. The PW column gives the value of precipitable water fitted to the asteroid spectrum, and is representative of its value over the course of the full night.

| Asteroid | Size | a | $p_v$ | family | Bus Class | Tholen Class | Density (cgs) |
|---|---|---|---|---|---|---|---|
| 1 Ceres | 952 | 2.765 | 0.09 | N | C | G | 2.13 ± 0.15 |
| 10 Hygiea | 407 | 3.1398 | 0.081 | Y | C | C | 2.19 ± 0.42 |
| 704 Interamnia | 307 | 3.0598 | 0.078 | N | B | F | 1.96 ± 0.28 |
| 52 Europa | 304 | 3.0108 | 0.057 | N | C | CF | 1.52 ± 0.39 |
| 31 Euphrosyne | 267 | 3.155 | 0.053 | Y | Cb | C | 1.18 ± 0.61 |
| 451 Patientia | 254 | 3.061 | 0.085 | N | Cb* | CU | 1.6 ± 0.80 |
| 88 Thisbe | 232 | 2.769 | 0.067 | N | B | CF | 3.44 ± 0.84 |
| 324 Bamberga | 221 | 2.686 | 0.050 | N | Cb* | CP | 1.52 ± 0.20 |

**Table 2:** Additional properties of the target asteroids. Bus Class taxonomic assignments are from (Bus and Binzel 2002) except * which are from the S3OS2. Densities are from Carry (2012).

| Date | Asteroid | 3-$\mu$m type | Region 2 Band Center (poly) | Region 2 Band Depth (poly) | Region 3 Band Center (poly) | Region 3 Band Depth (poly) | Region 2 Band Center (Gauss) | Region 2 Band Depth (Gauss) |
|---|---|---|---|---|---|---|---|---|
| 18 Sep 2006 | 1 Ceres | Ceres | 3.050 | 0.20 | 3.325 | 0.14 | 3.061 ± 0.002 | 0.20 |
| 23 Aug 2002 | 10 Hygiea | Themis | 3.040 | 0.10 | -- | | 3.040 ± 0.015 | 0.09 |
| 18 May 2005 | | Ceres? | 3.030 | 0.10 | 3.315 | 0.04 | 2.953 ± 0.017 | 0.10 |
| 8 Sep 2006 | | Themis | 3.070 | 0.08 | -- | | 3.058 ± 0.007 | 0.07 |
| 17 Sep 2006 | | Pallas? | 2.960 | 0.10 | -- | | 3.018 ± 0.014 | 0.10 |
| 12 Sep 2007 | | Themis | 3.080 | 0.13 | 3.360 | 0.08 | 3.095 ± 0.009 | 0.12 |
| 13 Sep 2007 | | Ceres | 3.055 | 0.13 | 3.355 | 0.07 | 3.048 ± 0.005 | 0.12 |
| 27 Jun 2012 | | Ceres | 3.045 | 0.12 | 3.335 | 0.06 | 3.055 ± 0.003 | 0.12 |
| 7 Jan 2008 | 24 Themis* | Themis | 3.115 | 0.13 | 3.360 | 0.07 | 3.102 ± 0.004 | 0.09 |
| 21 Sep 2005 | 31 Euphrosyne | Themis | 3.110 | 0.08 | 3.360 | 0.04 | 3.094 ± 0.010 | 0.09 |
| 28 Aug 2011 | | Themis | 3.120 | 0.06 | 3.380 | 0.02 | 3.112 ± 0.017 | 0.06 |
| 30 Jan 2012 | | Themis | 3.115 | 0.07 | 3.355 | 0.03 | 3.102 ± 0.010 | 0.06 |
| 8 Jan 2013 | | Themis | 3.075 | 0.08 | 3.375 | 0.02 | 3.084 ± 0.013 | 0.07 |
| 17 May 2006 | 52 Europa | Themis | 3.140 | 0.14 | -- | | 3.134 ± 0.0077 | 0.14 |
| 28 Jun 2012 | | Themis | 3.045 | 0.13 | -- | | 3.091 ± 0.016 | 0.13 |
| 27 Jun 2013 | | Themis? | 3.050 | 0.09 | -- | | 3.089 ± 0.013 | 0.10 |
| 21 Jul 2013 | | Themis | 3.090 | 0.07 | -- | | 3.094 ± 0.016 | 0.10 |
| 22 Jul 2009 | 88 Thisbe | Themis | 3.065 | 0.07 | -- | | 3.083 ± 0.013 | 0.09 |
| 7 Sep 2009 | | Themis? | 3.145 | 0.05 | -- | | 3.088 ± 0.083 | 0.05 |
| 20 Mar 2002 | 324 Bamberga | Themis? | 3.190 | 0.14 | -- | | 3.205 ± 0.011 | 0.14 |
| 30 Apr 2007 | | Themis | 3.140 | 0.08* | -- | | 3.145 ± 0.010 | 0.09 |
| 27 Jun 2012 | | Ceres? | 3.145 | 0.11* | -- | | 3.134 ± 0.012 | 0.07 |
| 3 Jul 2012 | | Themis | 3.115 | 0.12* | -- | | 3.126 ± 0.011 | 0.12 |
| 27 Jun 2013 | | Ceres? | 3.070 | 0.07 | 3.370 | 0.02 | 3.067 ± 0.009 | 0.07 |
| 21 Jul 2013 | | Pallas? | 2.965 | 0.04 | -- | | 3.040 ± 0.017 | 0.04 |
| 25 Aug 2013 | | Themis | 3.100 | 0.05 | -- | | 3.065 ± 0.015 | 0.06 |
| 29 Sep 2013 | | Ceres | 3.075 | 0.06 | 3.370 | 0.03 | 3.060 ± 0.004 | 0.06 |
| 30 Oct 2010 | 451 Patientia | Themis? | 3.045 | 0.13 | 3.280 | 0.11 | | |
| 28 Aug 2011 | | Ceres | 3.075 | 0.10 | 3.345 | 0.06 | 3.056 ± 0.010 | 0.06 |
| 17 Sep 2006 | 704 Interamnia | Pallas | < 2.95 | > 0.13 | 3.40** | 0.06 | 2.857 ± 0.031 | 0.12 |
| 12 Sep 2007 | | Ceres | 3.085 | 0.08 | -- | | 3.086 ± 0.016 | 0.06 |
| 29 Aug 2012 | | Pallas | < 2.95 | > 0.06 | 3.38 | 0.03 | 2.969 ± 0.015 | 0.07 |
| 1 Sep 2012 | | Ceres? | 3.045 | 0.07 | 3.40** | 0.05 | 3.025 ± 0.013 | 0.06 |
| 11 Dec 2013 | | Ceres? | 3.045 | 0.06 | 3.345 | 0.03 | 3.072 ± 0.027 | 0.10 |

**Table 3:** Fitted band centers and depths using the polynomial and Gaussian fits, and the 3-$\mu$m type assigned by inspection. Polynomial band center and depth uncertainties are estimated to be 0.0025μm and ~0.005, respectively. *A different tie point at long wavelengths is used for continuum calculation to avoid a spectral artifact near 3.7 $\mu$m. **The polynomial fit suggests a band center > 3.40 $\mu$m, but the polynomial fit is not valid at wavelengths > 3.40 $\mu$m

*Figures*

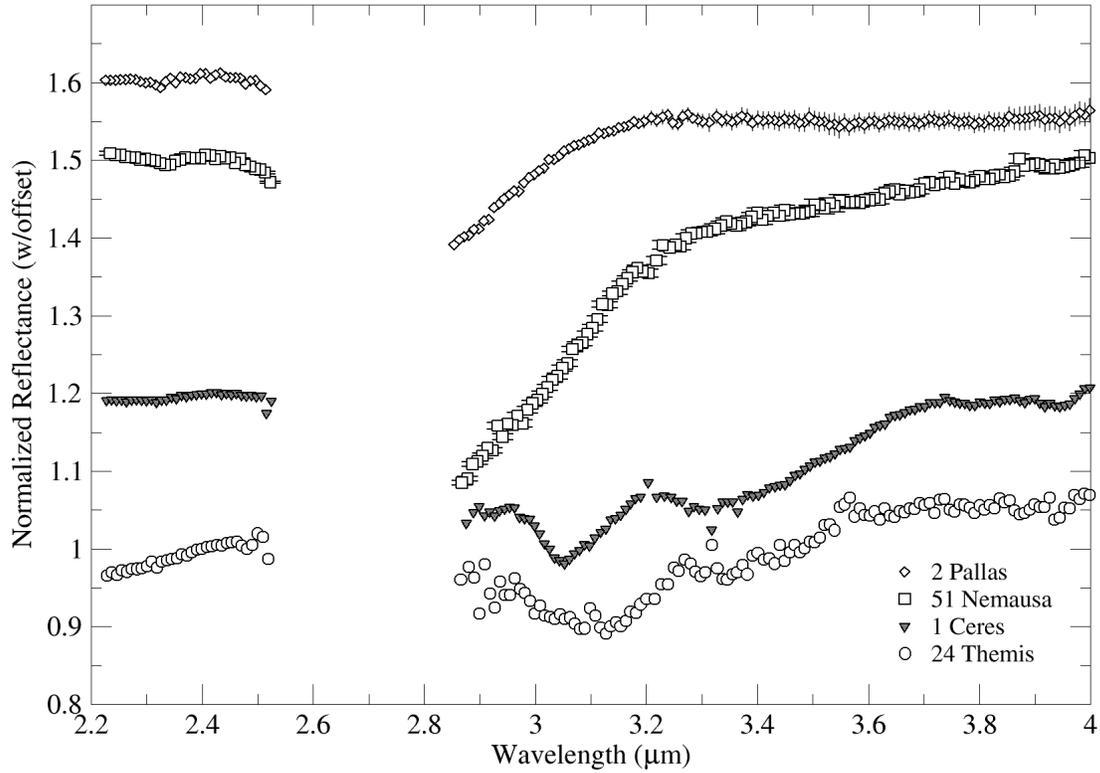

**Figure 1:** The major band shapes in the low-albedo population discussed in this paper: "Pallas types" (51 Nemausa from Rivkin et al. 2015b and 2 Pallas from Rivkin et al. 2014b) with a band attributed to phyllosilicates and seen in the meteorite collection; "Ceres types" (1 Ceres from this work, with a sub-Earth latitude of +2° on an observation date of 18 September 2006) attributed to ammoniated phyllosilicates, and carbonates and not seen in the meteorite collection; "Themis types" (24 Themis from Rivkin and Emery 2010) attributed to ice frost and organic materials, also not seen in the meteorite collection.

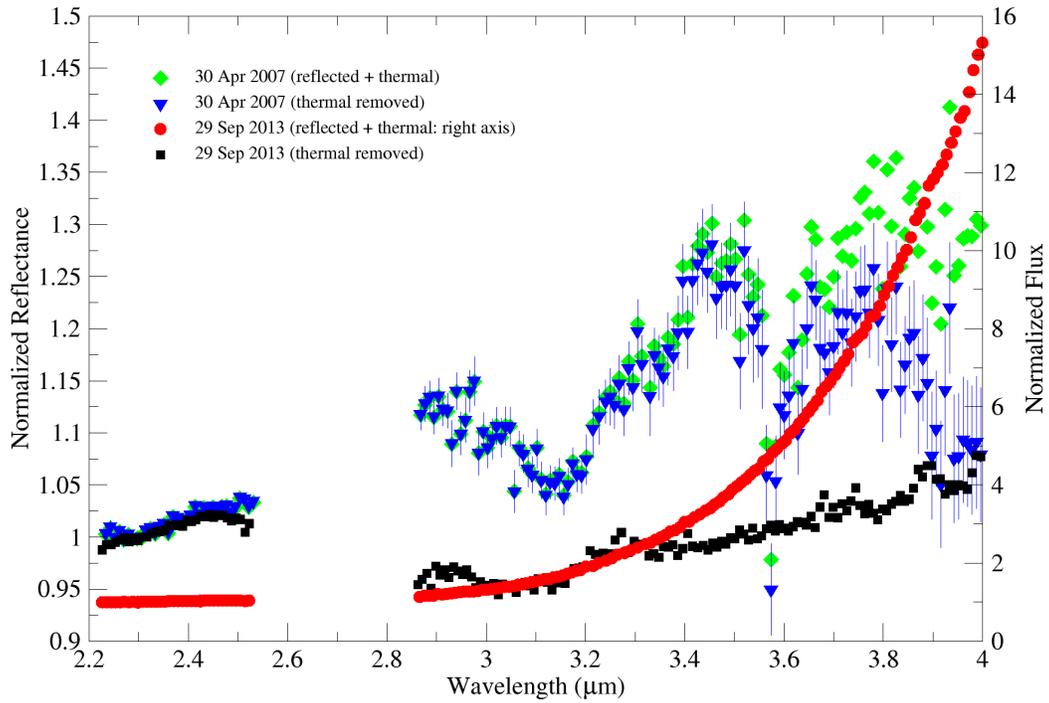

**Figure 2:** Asteroid 324 Bamberga on two dates, with thermal emission and after thermal flux removal. On 30 April 2007 it was 3.58 AU from the Sun, with minimal thermal emission even at the longest wavelengths. On 29 September 2013 it was 1.79 AU from the Sun, and thermal emission dominates its spectrum beyond 3.2 $\mu$m. Note that the 29 September reflected + thermal spectrum uses the right axis rather than the left axis.

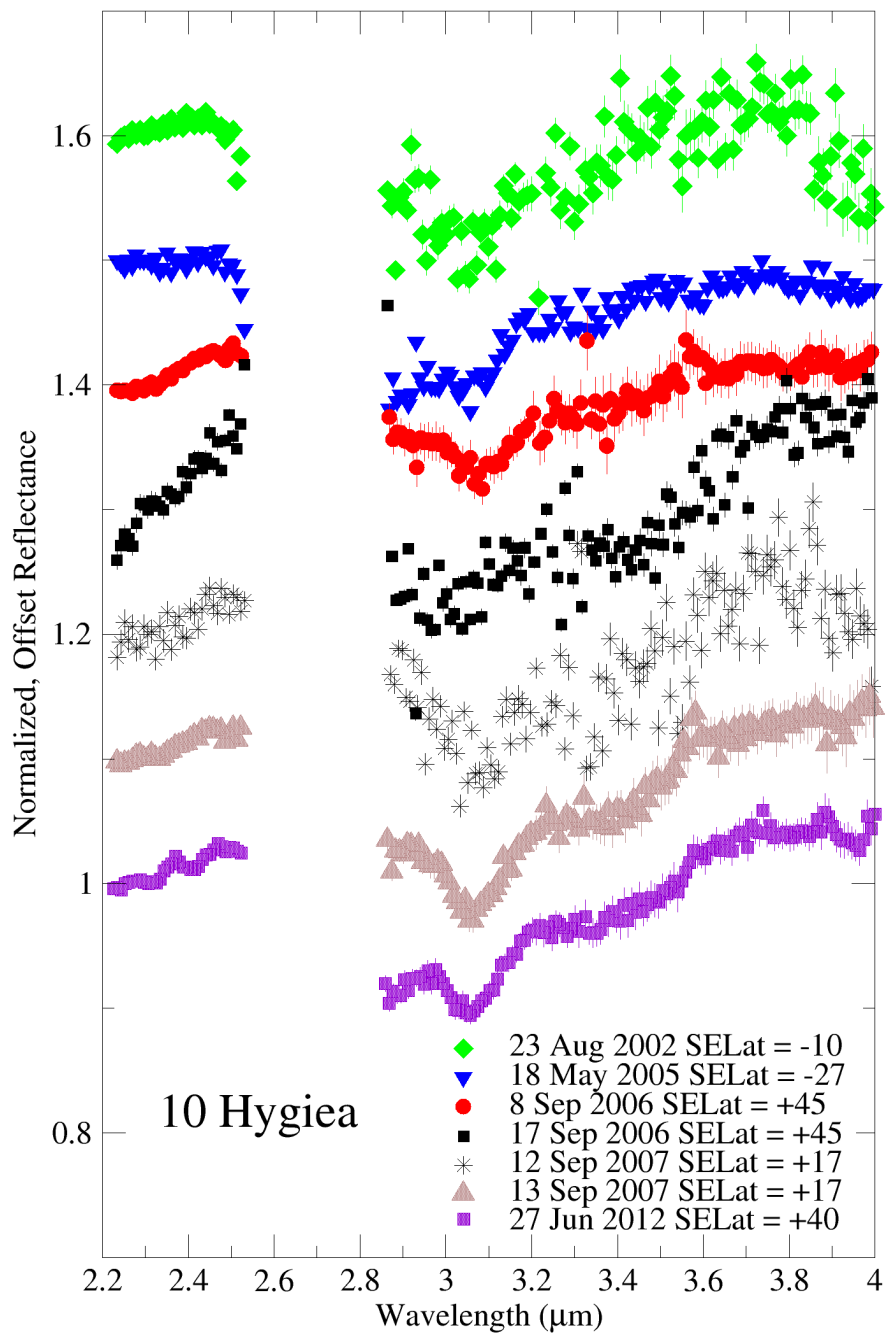

**Figure 3:** Spectra of 10 Hygiea, offset for clarity.

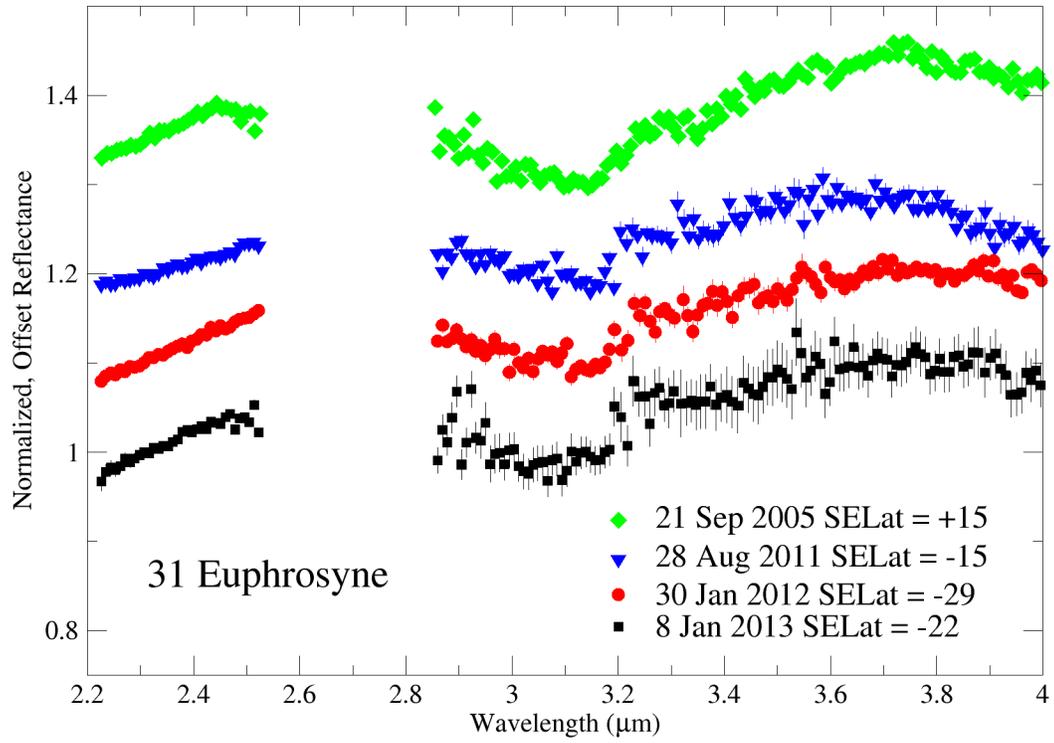

**Figure 4:** Spectra of 31 Euphrosyne, offset for clarity.

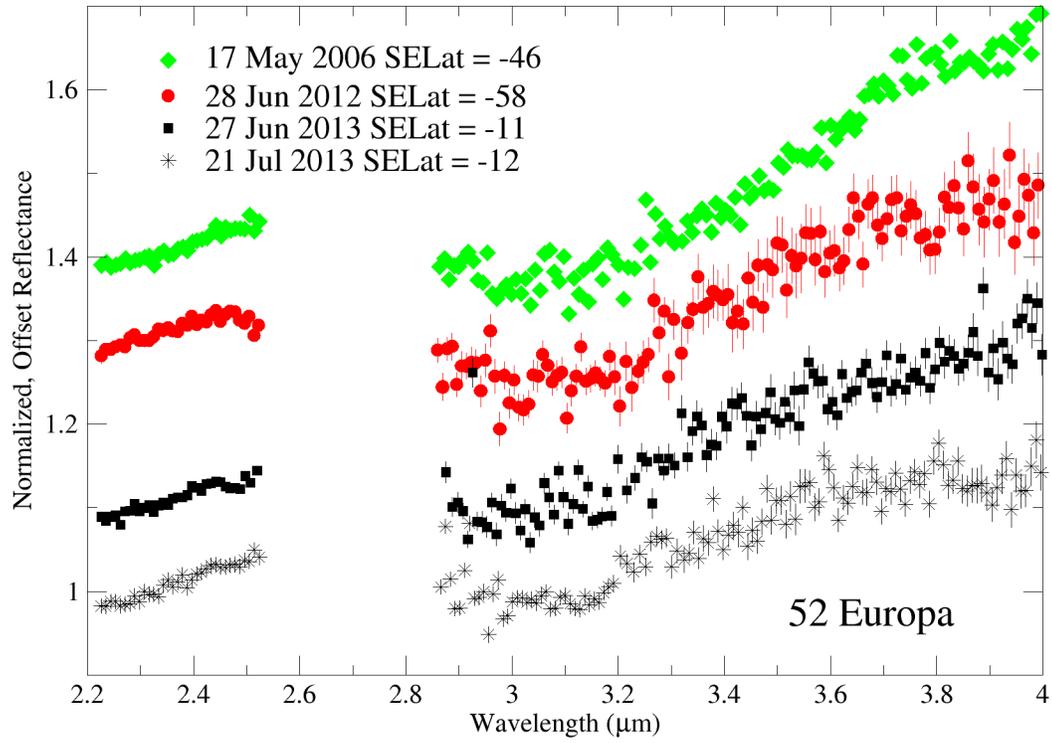

**Figure 5:** Spectra of 52 Europa, offset for clarity

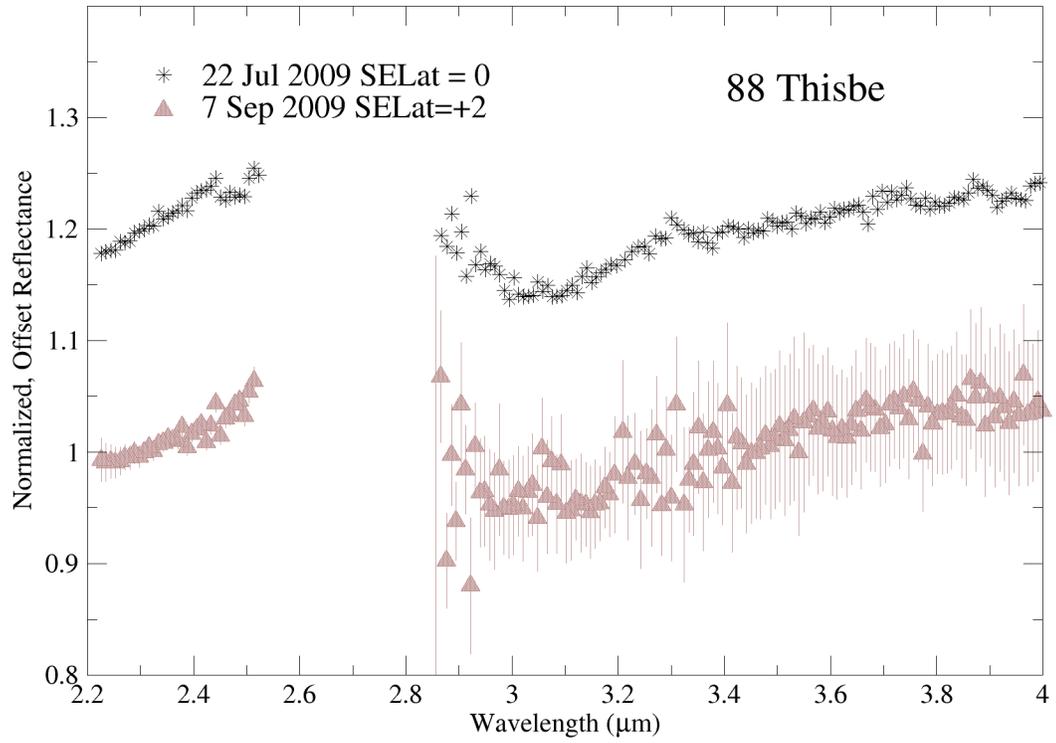

**Figure 6:** Spectra of 88 Thisbe, offset for clarity

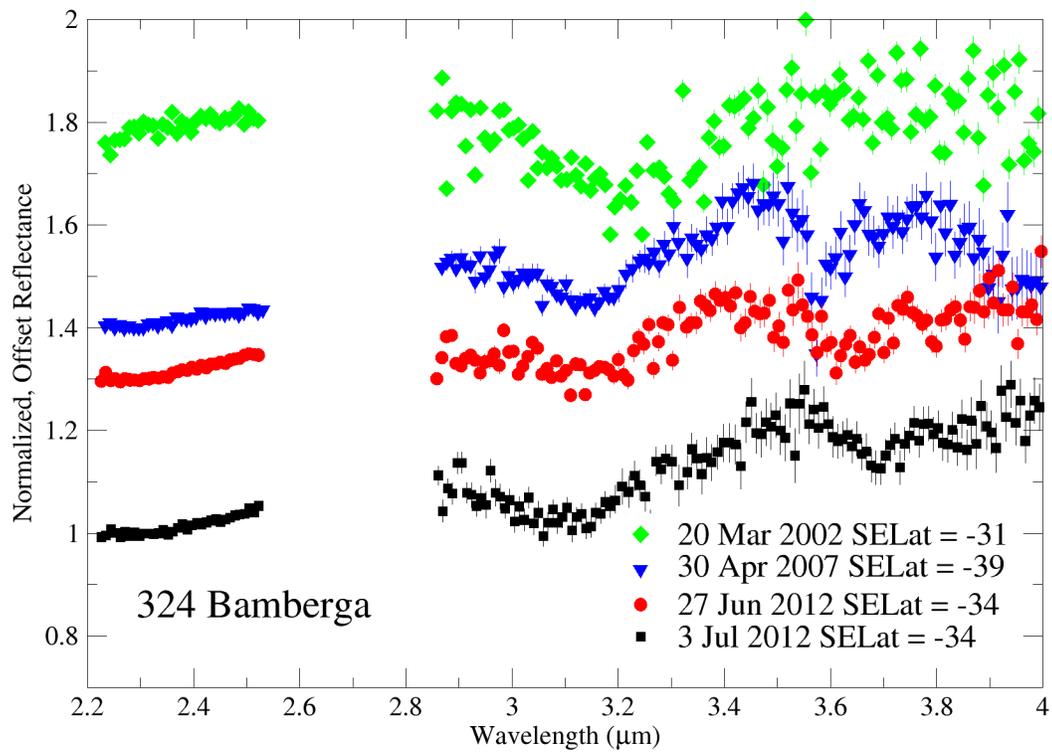

**Figure 7:** New spectra of 324 Bamberga prior to 2013, offset for clarity. These spectra all are of Bamberga's southern hemisphere. The apparent feature near 3.5-3.7 µm is interpreted as an artifact due to an overlap between grating orders.

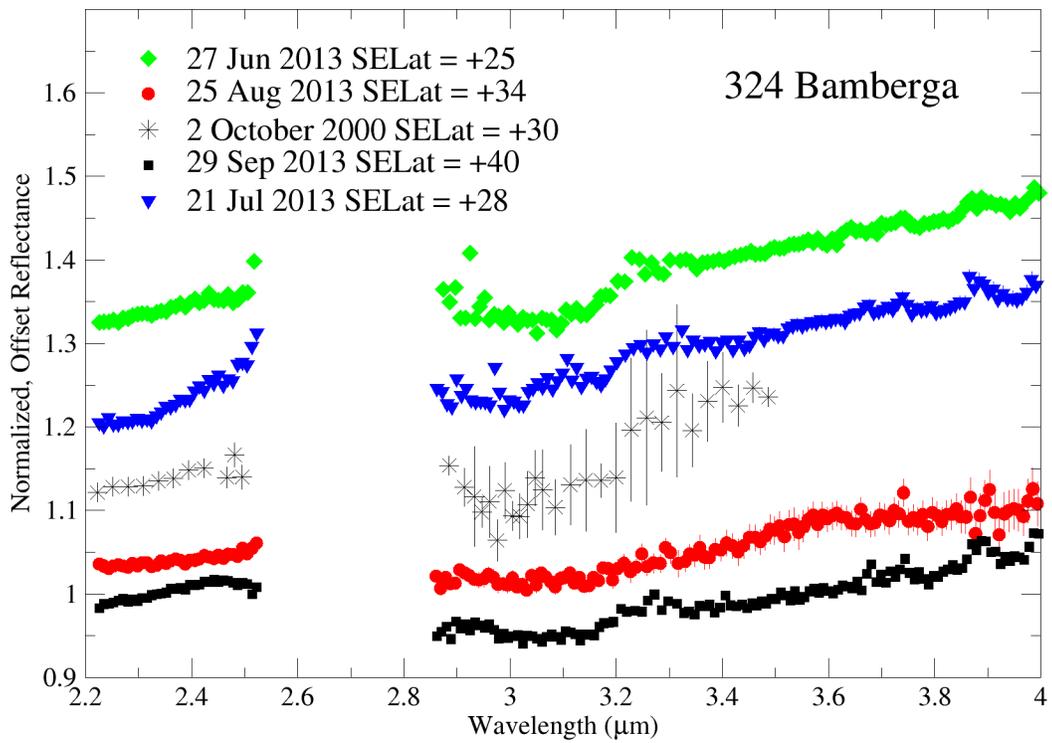

**Figure 8:** New spectra of 324 Bamberga from 2013 onward, offset for clarity. All of these spectra were taken of Bamberga's northern hemisphere. In addition to the new spectra is a spectrum from Rivkin et al. (2003), also of Bamberga's northern hemisphere.

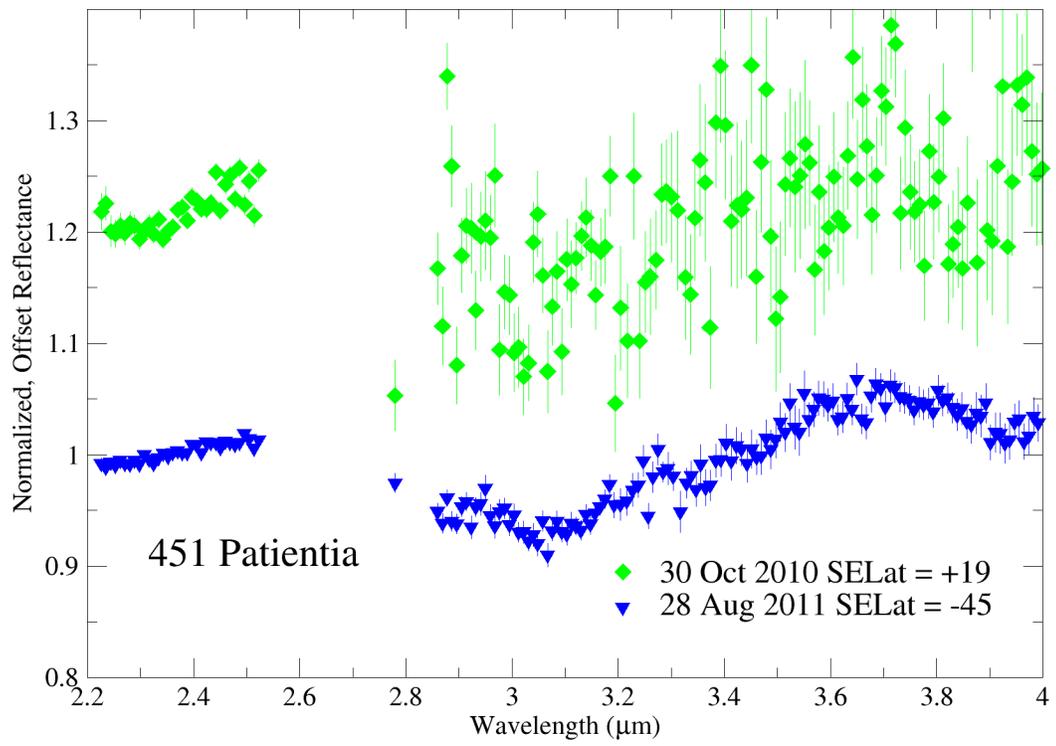

**Figure 9:** Spectra of 451 Patientia, offset for clarity.

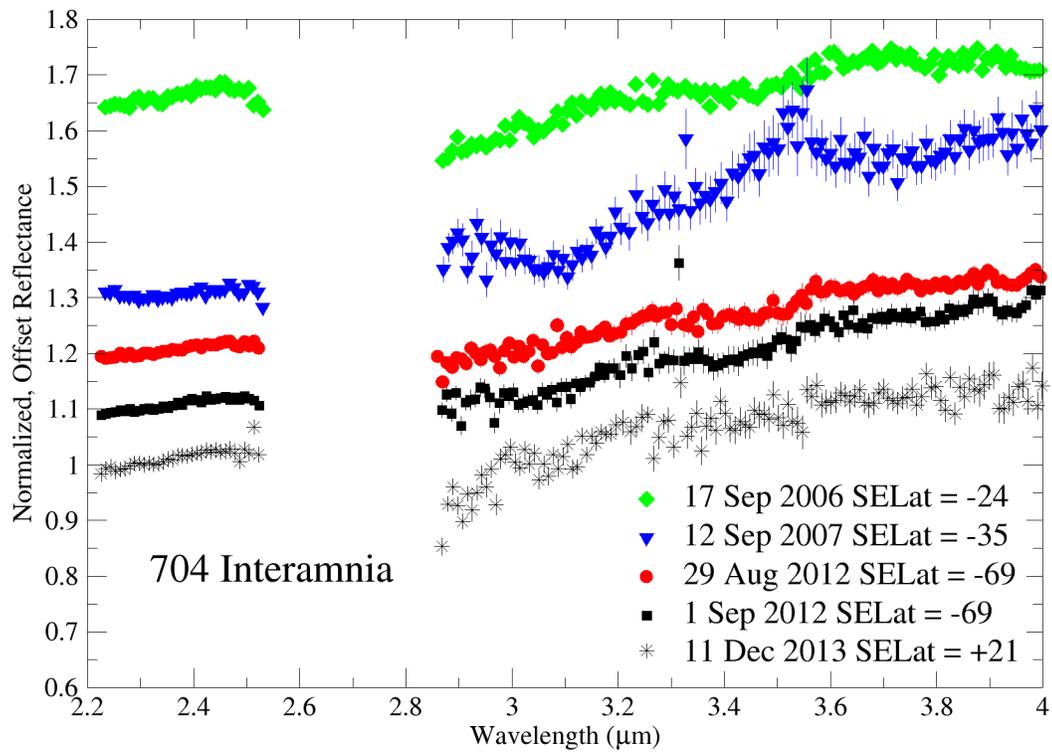

**Figure 10:** Spectra of 704 Interamnia, offset for clarity. The positive feature near 3.55 $\mu$m is interpreted as an artifact similar to those discussed in Figure 8.

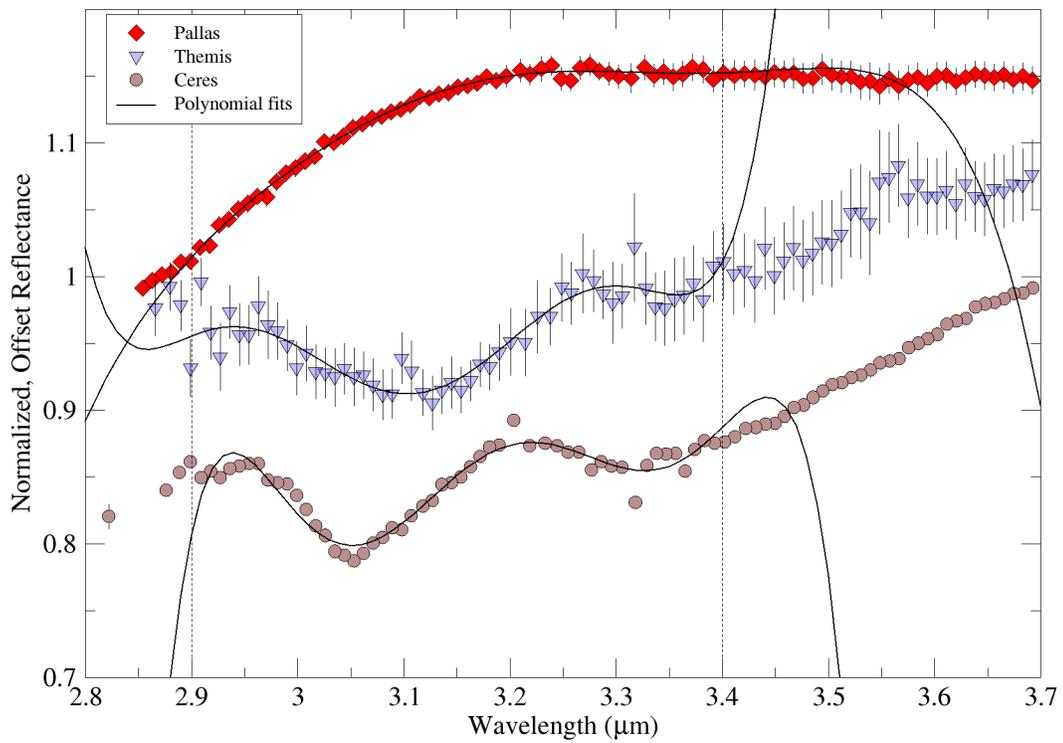

**Figure 11:** The spectra of type objects for the 3-μm band shapes are shown here offset from one another, with the 6th-order polynomial fits to each. The vertical dotted lines show the range over which the polynomials were fit and are thus valid. In each case the band shape is well fit, with only small deviations from the real data within the region of validity.

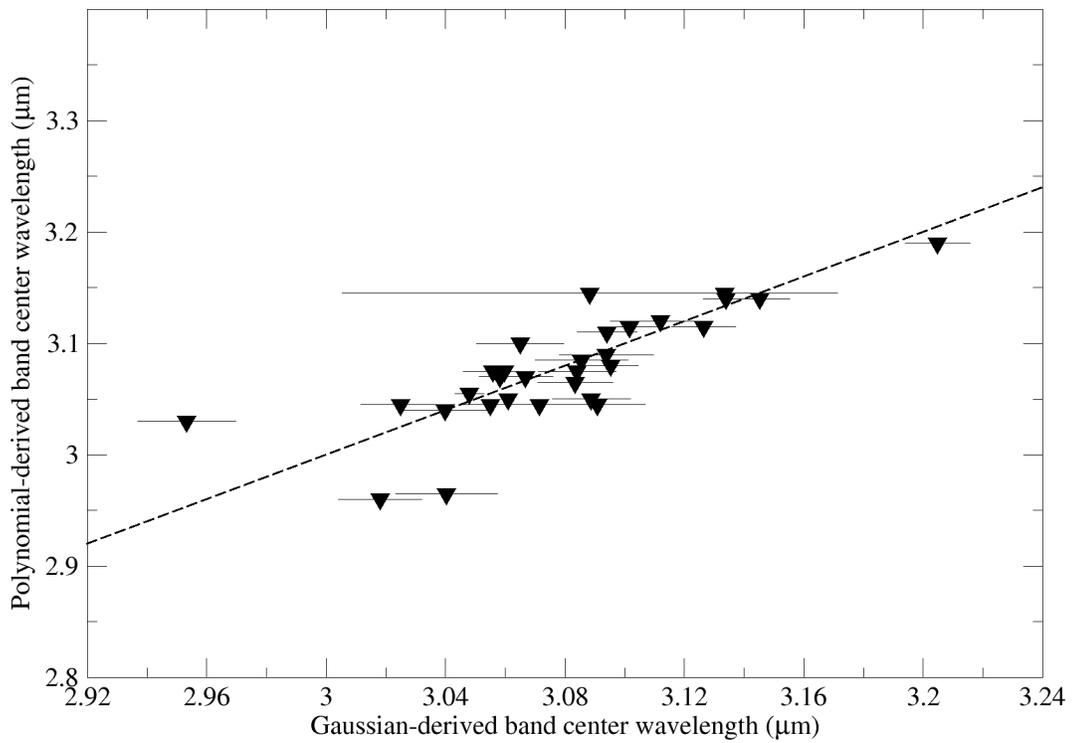

**Figure 12:** Two techniques were used to determine band centers: A sixth-order polynomial fit and a Gaussian fit. The two techniques are in good agreement: the dashed line indicates where band centers with exactly the same value in both techniques would fall.

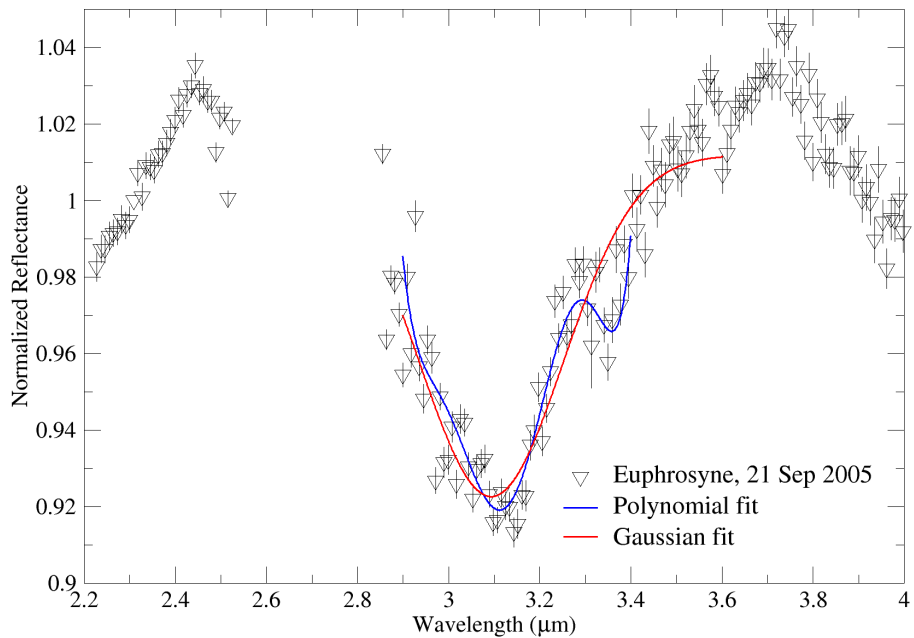

**Figure 13:** As noted, the spectra in this work were fitted with both a polynomial and Gaussian. This spectrum of Euphrosyne shows the differences and similarities in their results. Both find a band near 3.1 $\mu$m with a depth of ~9%, though they differ from one another by ~0.015 $\mu$m in band center and ~1% in depth. The polynomial approach is more sensitive to asymmetries, and fits a band in Region 3 centered at 3.36 $\mu$m. Because only a single Gaussian is used in the Gaussian fits, it does not identify such a band. It is not clear in this spectrum whether the 3.36-$\mu$m band is real or due to noise, though several of the objects in the asteroid sample have more obvious absorptions near that wavelength.

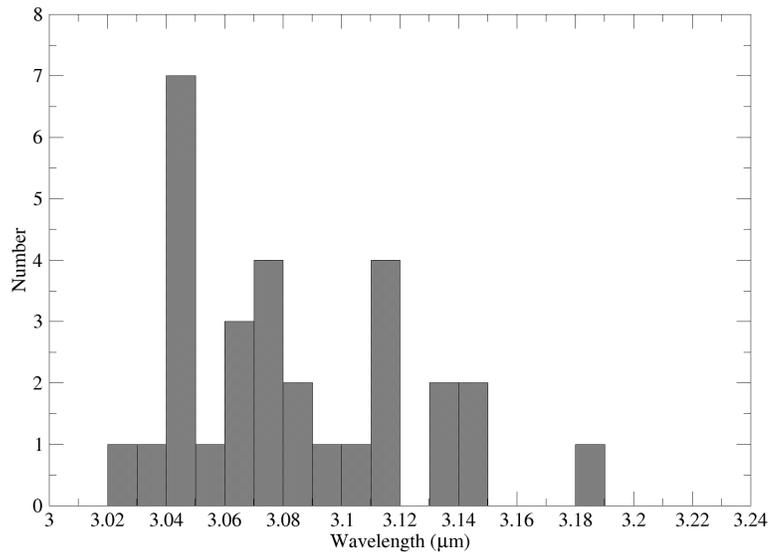

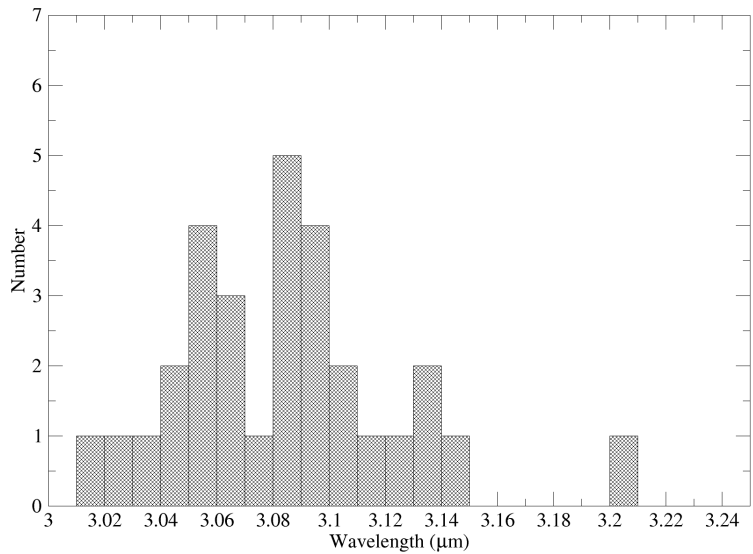

**Figure 14:** Histograms of Band 1 centers from the polynomial fit (top) and Gaussian fit (bottom). Both show clusters near the band center for Ceres (3.06 $\mu$m) and Themis (3.11 $\mu$m), but neither shows a hiatus that appears to be appropriate for use in dividing those two groups from one another.

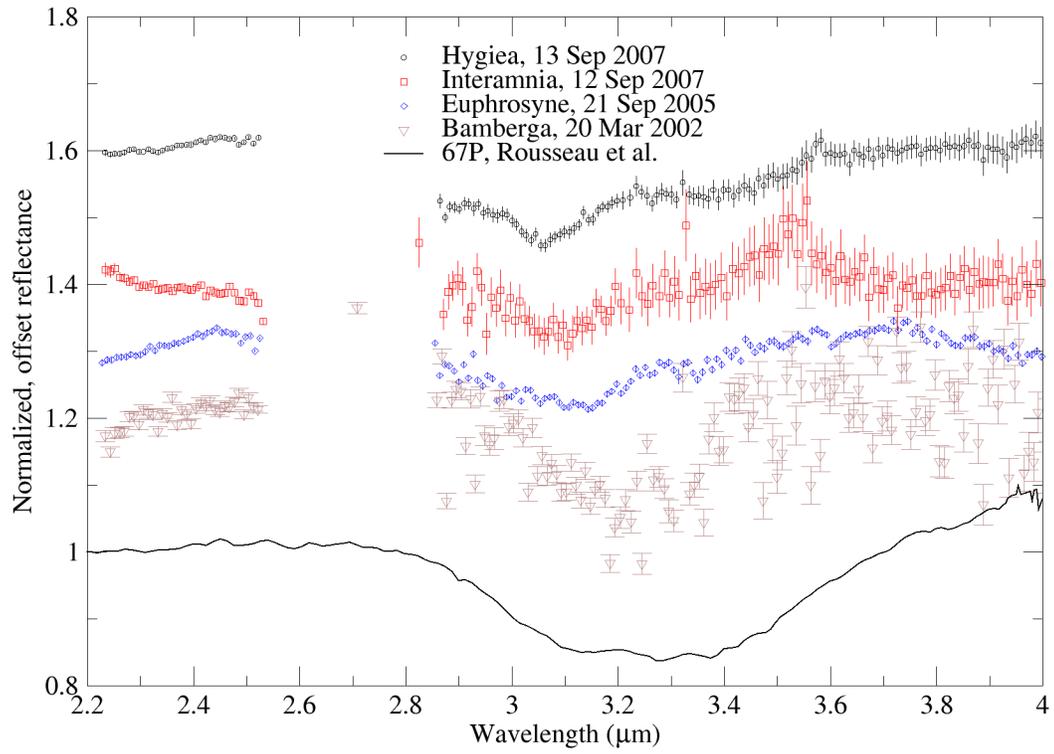

**Figure 15:** The wavelengths of Band 1 centers in the Ceres- and Themis-type asteroids range from shortward of 3.05 $\mu$m to 3.20 $\mu$m, close to the band center wavelength seen on 67P (Rousseau et al. 2018). It is possible that the minerals creating the broad band on 67P are also present on these asteroids.

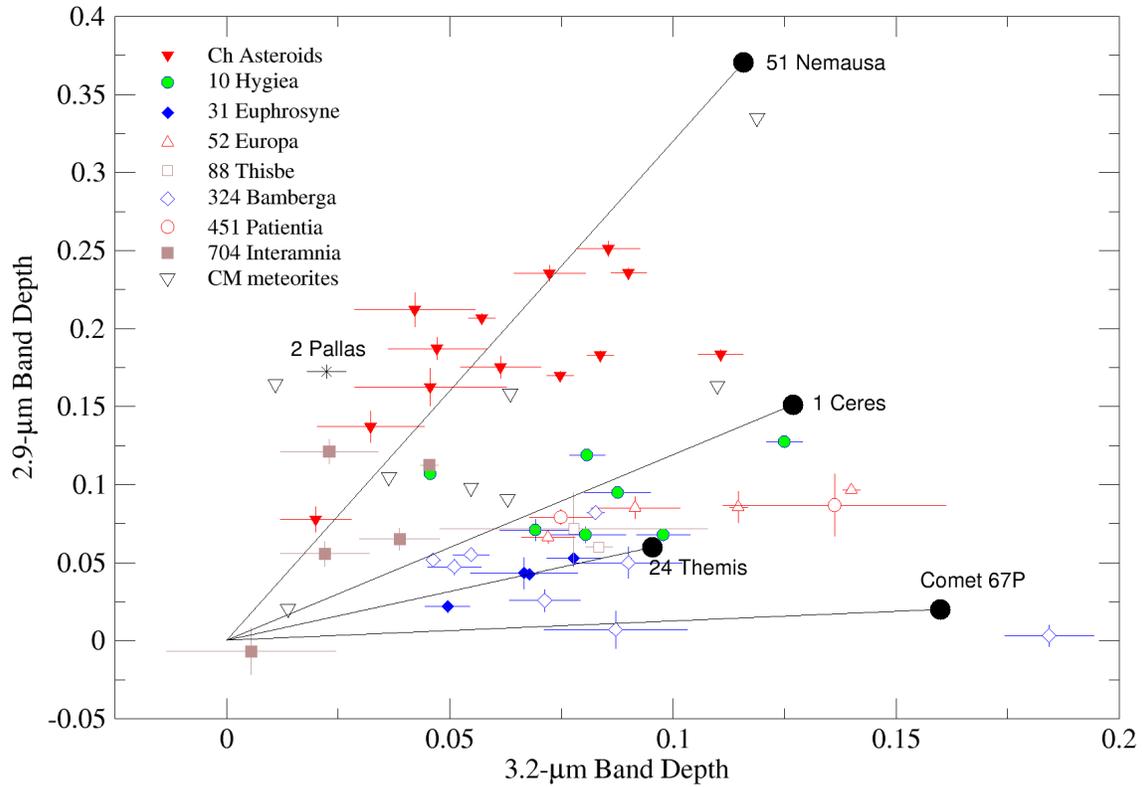

**Figure 16:** On a band depth-band depth chart, asteroids with different 3-µm band shapes fall in different areas. The Ch asteroids and the CM meteorites fall along or close to a trend connecting the origin with Nemausa, consistent with their classification as Pallas types. The majority of spectra listed in Table 3 cluster near Themis or between Themis and the line connecting Ceres with the origin. Two spectra of Bamberga fall near the line connecting Comet 67P to the origin, consistent with the suggestion from the previous figure that these objects may have similar surface compositions.

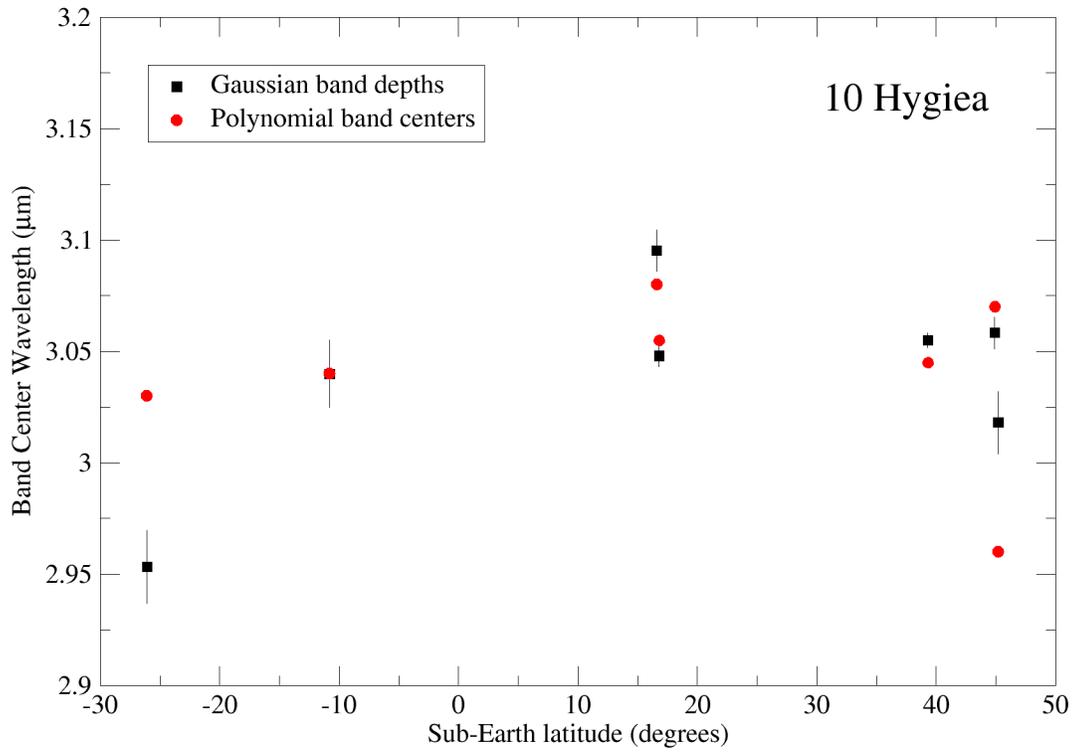

**Figure 17:** Band centers in Region 1/Region 2 vs. sub-Earth latitude on Hygiea. While band centers vary, no correlation with latitude is seen.

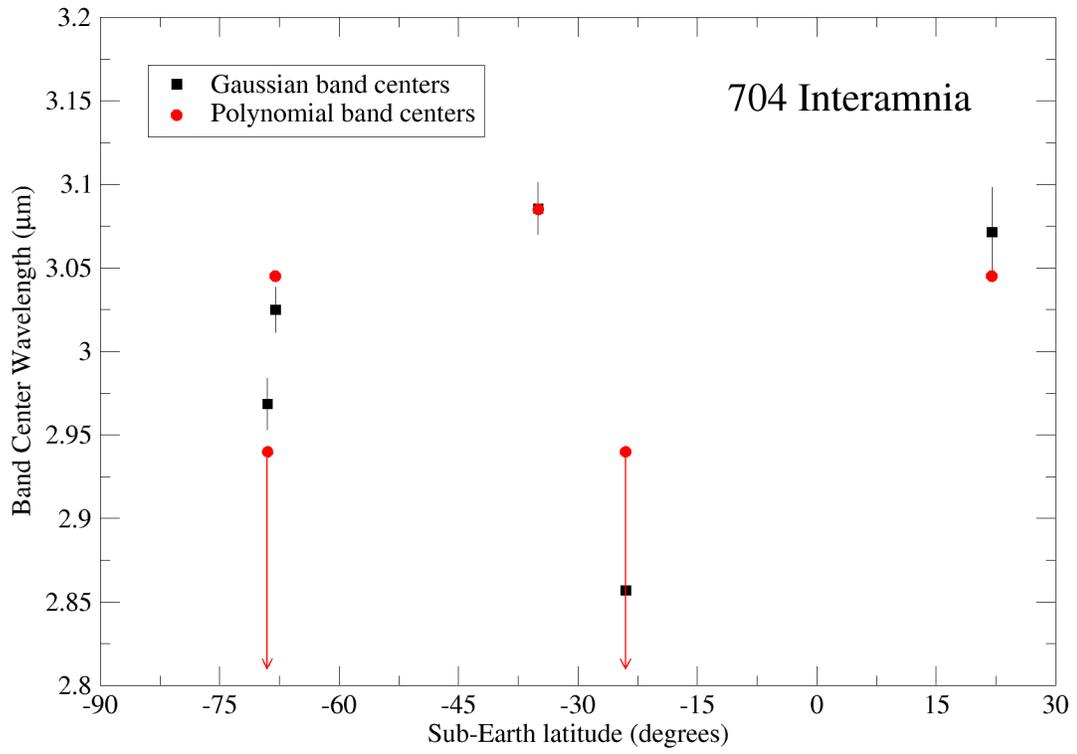

**Figure 18:** Same as previous figure, but for Interamnia. Arrows show long-wavelength limits of band centers that are not observable due to atmospheric opacity. Those band centers are likely between 2.7-2.8 $\mu$m.

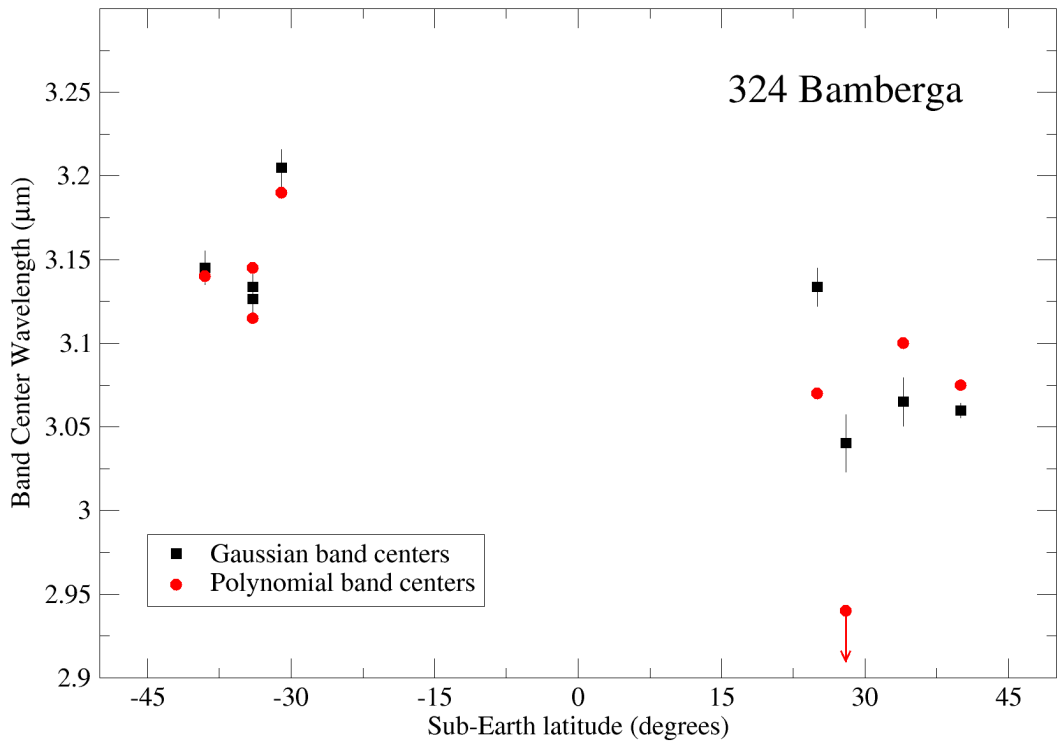

**Figure 19:** Same as previous figure, but for Bamberga. The band centers are systematically at longer wavelengths in Bamberga's southern hemisphere compared to its northern hemisphere.

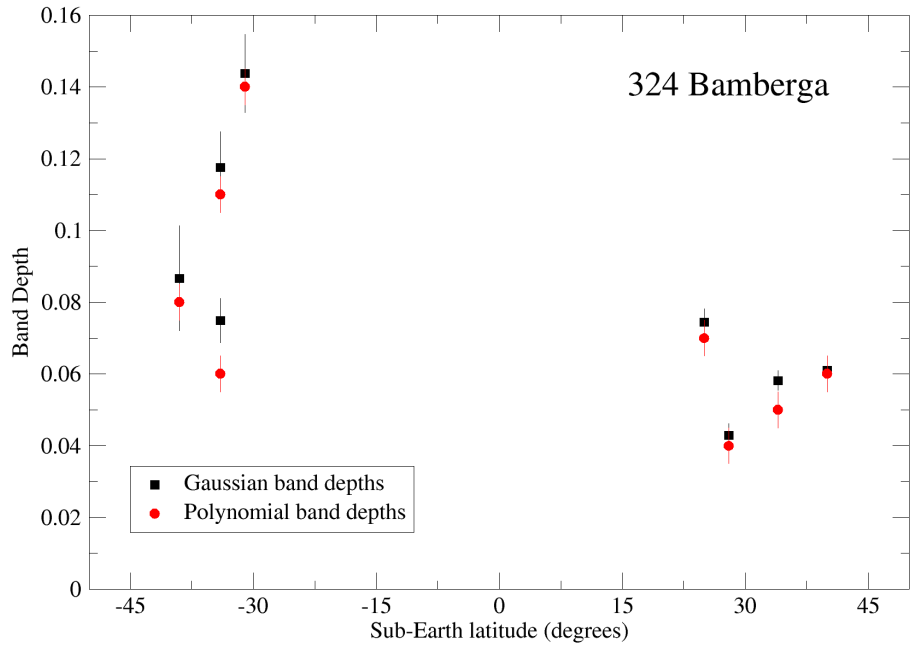

**Figure 20:** Same as the previous figure, but for band depths. The fits show deeper band depths in the southern hemisphere than the northern hemisphere, especially the Gaussian fits.